\documentclass[twocolumn]{aastex62}
\pdfoutput=1 
\usepackage{amsmath,amstext}
\usepackage[T1]{fontenc}
\usepackage{apjfonts} 
\usepackage[figure,figure*]{hypcap}
\usepackage{graphicx}
\usepackage{rotating}

\shorttitle{The Stellar CME-flare relation}
\shortauthors{S.P. Moschou et al.}

\begin{document}

\title{The Stellar CME-flare relation: What do historic observations reveal?}

\author{Sofia-Paraskevi Moschou}
\author{Jeremy J. Drake}
\affiliation{Center for Astrophysics | Harvard \& Smithsonian, 60 Garden Street, Cambridge MA 02138, USA}

\author{Ofer Cohen}
\affiliation{Lowell Center for Space Science and Technology, University of Massachusetts, Lowell, Massachusetts, USA}

\author{Juli\'an D. Alvarado-G\'omez}
\affiliation{Center for Astrophysics | Harvard \& Smithsonian, 60 Garden Street, Cambridge MA 02138, USA}

\author{Cecilia Garraffo}
\affiliation{Center for Astrophysics | Harvard \& Smithsonian, 60 Garden Street, Cambridge MA 02138, USA}
\affiliation{IACS, Harvard University, 33 Oxford Street Cambridge, MA 02138, USA}

\author{Federico Fraschetti}
\affiliation{Center for Astrophysics | Harvard \& Smithsonian, 60 Garden Street, Cambridge MA 02138, USA}
\affiliation{Dept. of Planetary Sciences-Lunar and Planetary Laboratory, University of Arizona, Tucson, AZ, 85721, USA}

\begin{abstract}
Solar CMEs and flares have a statistically well defined relation, with more energetic X-ray flares corresponding to faster and more massive CMEs. How this relation extends to more magnetically active stars is a subject of open research. Here, we study the most probable stellar CME candidates associated with flares captured in the literature to date, all of which were observed on magnetically active stars. We use a simple CME model to derive masses and kinetic energies from observed quantities, and transform associated flare data to the GOES 1--8~\AA\ band. 
Derived CME masses range from $\sim 10^{15}$ to $10^{22}$~g. Associated flare X-ray energies range from $10^{31}$ to $10^{37}$~erg.
Stellar CME masses as a function of associated flare energy generally lie along or below the extrapolated mean for solar events. In contrast, CME kinetic energies lie below the analogous solar extrapolation by roughly two orders of magnitude, indicating approximate parity between flare X-ray and CME kinetic energies. These results suggest that the CMEs associated with very energetic flares on active stars are more limited in terms of the ejecta velocity than the ejecta mass, possibly because of the restraining influence of strong overlying magnetic fields and stellar wind drag. 
Lower CME kinetic energies and velocities present a more optimistic scenario for the effects of CME impacts on exoplanets in close proximity to active stellar hosts.
\end{abstract}


\section{Introduction}
\label{S:intro}
For decades, since the first space-based coronal mass ejection (CME) observations in the 1970s \citep{Tousey:73}, the Sun has been the only star that allowed for direct CME observation. Recently, with the discovery of multiple exoplanetary systems, there is an increasing scientific interest to determine the effects of stellar activity on planetary atmospheres and habitability \citep[e.g.][]{Khodachenko.etal:07,Khodachenko.etal:07b}.  These efforts are similar in some ways to what has been done up to now in the field of space weather \citep{Kahler.etal:01,Zhang.etal:07,Webb.etal:09,Yashiro2009,Cane.etal:10,Vourlidas2011,Cliver.Dietrich:13,Reames:13,Gopalswamy:16}. While space weather in the solar environment is comparatively {better understood}, the vast majority of the stars in our Galaxy are red dwarfs and only a small percentage are Sun-like stars. The magnetic behaviour of stars quite different from the Sun remains uncertain in detail, and it has become imperative that the role of stellar CMEs is assessed in this context on other stars \citep[e.g.][]{Kay.etal:16}.

Solar CMEs and flares are more tightly associated with each other with increasing flaring energy \cite[e.g.][]{Yashiro2009,Aarnio2011SoPh} with the association reaching a '1-1' ratio for high energies. Solar flares are classified morphologically as either compact with a small number of magnetic loops flaring up for a few minutes and ''two-ribbon'' flares that unwind over longer time-scales on the order of hours {\citep[see, e.g.][]{Pallavicini.etal:77,Shibata.Magara:11}}. Two-ribbon flares are associated with an arcade group with complex topology and footpoints that form two parallel chromospheric ribbons visible in $H_\alpha$. Very large solar flares belong in the two-ribbon class, which involves a continuous reconnection starting at the top of the magnetic arcade system and propagating upward toward loops positioned on top of each other flaring serially. 

A plethora of stellar flares have been observed in radio, optical, UV and X-ray wavelengths in active and Sun-like stars and in both single and binary star systems \citep[e.g.][]{Osten:05,Huenemoerder.etal:10,Kretzschmar:11,Notsu.etal:16,Crosley.Osten:18a}. All classes of late-type stars are known to flare in soft X-rays \citep[e.g.][]{Schmitt:94}. \citet{Pandey.Singh:08} showed that a) late-type G-K dwarfs flare frequently, b) their flares resemble the two-ribbon solar ones, and c) even though they are as energetic as M dwarf flares, they are energetically weaker than flares from pre-main-sequence, giant and dMe stars. Giant flares from binary systems and young stars with a disk could result from magnetic coupling between the binary members and the young star and its disk, respectively \citep{Graffagnino.etal:95,Grosso.etal:97,Tsuboi.etal:00}.
As argued in \cite{Drake.etal:13}, strong winds of active stars are potentially dominated by CMEs, as a result of their extreme flaring activity, with great implications for the energy budget of the system \citep[see also][]{Aarnio.etal:12,Osten:15}. Later on \citet{Odert.etal:17} developed a model for estimating mass-loss rates due to CMEs in other stars using the solar flare-CME relations and stellar flaring rates. The authors suggest that solar extrapolations present limitations in their applicability to the young-star regime  as they reached CME-driven mass-loss rates higher than the total observationally determined values. In a different approach \citet{Cranmer:17} exploited correlations between solar surface-averaged magnetic flux values and mean kinetic energy flux in CMEs and the wind to predict CME and wind mass-loss rates in other stars. The main results therein was that the mass-loss rate for stars younger than 1 Gyr is dominated by CMEs. More recently, \citet{Vida.etal:19} presented a large statistical analysis of $\sim$500 stellar events with line asymmetries in Doppler-shift observations. They measured speeds of the order of 100 \,--\, 300 km s$^{-1}$ and masses in the order of $10^{15}-10^{18}$~g and confirmed that cooler stars appear more chromospherically active.

\citet{Pallavicini.etal:90} published a thorough stellar flare survey using \textit{EXOSAT} (The European X-ray Observatory SATellite) which brought to light two stellar flare classes. The first class involved impulsive flares that resemble compact solar flares, while the second class involved flares with longer decay times, which are similar to the two-ribbon solar flare category. The impulsive flares have a short (less than an hour) duration, emit a total X-ray energy in the order of $L_X=10^{30}  \ \mathrm{erg\ s^{-1}}$, and are believed to involve a single loop only. The flares with longer decay are two orders of magnitude more energetic $L_X=10^{32}\ \mathrm{erg\ s^{-1}}$, last more than an hour, and involve an arcade forming group of loops. As \cite{Pandey.Singh:12} emphasized, even though there are several similarities between solar and stellar flares it is difficult to draw a direct parallel as the latter ones involve orders of magnitude larger energies \citep{Gudel.Naze:09}.

It is essential to understand the characteristics of stellar CMEs in order to evaluate the habitability of an exoplanet. Active stars are observed to flare so frequently \citep[e.g.][]{Kashyap.etal:02,Huenemoerder.etal:10} that their light curves can often times be approximated by a superposition of flares \citep[e.g.][]{Audard.etal:00,Caramazza.etal:07}. If a high association rate between stellar CMEs and flares is in place for active stars then exoplanets orbiting them would face frequent interactions with transients. This can lead to very high depletion rates as shown in \citet{Cherenkov.etal:17}, but the planetary magnetic field might be able to shield the planetary atmosphere \citep[see, e.g.][]{Cohen.etal:11}.

With current instrumentation we are unable to directly observe CMEs even in the closest stars and thus one has to turn to indirect observational evidence for signatures of this eruptive phenomenon. The absorption of emission coming from underlying stellar atmospheric layers by the CME volume is one indirect method especially useful for large stellar mass eruptions rather than solar-like ones. Absorption is observed in solar erupting filaments as well {\citep[e.g.][]{Subramanian.Dere:01,Kundu.etal:04,Jiang.etal:06,Vemareddy.etal:12,Gosain.etal:16,Chandra.Filippov:17}}, but the CME material does not suffice to cause significant absorption by itself. If the CME\,--\,flare association holds for much more energetic events on other stars, we expect considerably more massive CMEs for active stars and their masses could then provide a valuable observational tool through absorption.

In Section~\ref{sec:analysis}, we introduce and briefly explain the Doppler shift and absorption methods currently available for CME tracing on stars other than the Sun. Then, we present all the known stellar CME candidates identified to date, and in Section ~\ref{sec:events} estimate their mass and kinetic energy and place them in the X-ray fluence \,--\, CME characteristic property diagram. In Section~\ref{sec:results}, we present our results and in Section~\ref{sec:discussion} we discuss sources of discrepancy, other proposed CME detection techniques and mention future missions and computational models that will contribute in the field in the near future. We wrap up this paper with our conclusions in Section~\ref{S:conclusions}.

\section{Analysis} \label{sec:analysis}

\subsection{Tracking methods}
Even though stellar flares and super-flares (with energies $>10^{33}~\mathrm{erg}$, see e.g. \citet{Notsu.etal:16}) are routinely observed in Sun-like and more active stars over a wide range of wavelengths \textcolor{blue}{---}ranging from radio to X-rays\textcolor{blue}{---}, the direct imaging of stellar CMEs is a difficult, if not impossible, task with current instrumentation. For that reason, several observational proxy methods have been proposed to indirectly provide evidence for CME occurrence in other stars \citep[see e.g.][]{Osten.Wolk:17}. However, only a handful of possible CME events have been captured for each of two techniques, namely X-ray continuous absorption \citep[see][]{Moschou:17} and Doppler shifts in UV wavelengths \citep{Vida.etal:16}.

As explained in \citet{Leitzinger.etal:14}, several suspected CMEs have been the focus of observational studies where X-ray absorption in association with energetic flares was seen \citep{Haisch.etal:83,Ottmann.Schmitt:96,Tsuboi.etal:98,Favata1999,Franciosini.etal:01,Pandey.Singh:12}, as well as from flare-associated Doppler shifts in Balmer lines \citep{Houdebine.etal:90,Guenther.Emerson:97,Bond.etal:01,Fuhrmeister.Schmitt:04,Leitzinger.etal:11,Vida.etal:16}. In the next Section, we will present all the CME candidates known from published studies so far based on those two methods.

Continuous X-ray absorption during a stellar flaring event can be used to infer the kinematic characteristics of the obscuring material, e.g. due to a CME {or a prominence.} The best representative of {a CME candidate observed through X-ray absorption} is the 1997 August 30 Algol event. This was a very large X-ray flare analyzed by \cite{Favata1999} and the parameters of the associated potential CME were derived by \cite{Moschou:17}. {The reason that a CME is the most probable scenario for the August 1997 Algol event is that there was sharp increase, by almost two orders of magnitude, in the column density and the continuous absorption in X-rays gradually decayed with an inverse square law with time. As discussed extensively in \citet{Moschou:17}, this temporal variation, combined with the lack of rotation modulation of the X-ray signal, is consistent with a CME expanding in a self-similar manner away from the stellar surface. Here we examine CME candidates that show a clear column density decay over periods of time of several ks and treat as prominences events with no substantial temporal decay, see Fig.~\ref{fig:xray}.} Wherever possible, we will follow a similar analysis technique here {as in \citet{Moschou:17}} to analyze all the suspected CMEs inferred through X-ray absorption. {It must be noted, however, that it is very difficult to exclude the prominence scenario even for historic CME candidates with a column density decrease without resolving the stellar surface. A prominence consisting of dense cool chromospheric material could undergo complex cooling and heating processes \citep[e.g.][]{Moschou.etal:15} which affect its ionization degree, and as a result the prominence could fade when observed in a specific passband~\citep[e.g.][]{Ballester.etal:18}.}

The observational method based on blue-shifted spectral lines involves larger uncertainties, because projection effects are difficult to disentangle \citep[e.g.][]{Leitzinger.etal:11}. The inferred velocities in several studies lie in the local chromospheric plasma flow range, i.e.\ a few tens up to about a couple of hundred $\mathrm{km\ s^{-1}}$ \citep{Bond.etal:01,Fuhrmeister.Schmitt:04,Leitzinger.etal:11}, making it difficult to discriminate them from other events such as chromospheric brightenings \citep{Kirk.etal:17} or chromospheric evaporation \citep{Teriaca.etal:03,Gupta.etal:18}. \citet{Leitzinger.etal:14} concluded that the CME strength, translated in terms of CME mass or flux is the most important parameter that controls the efficiency of the Doppler-shift method used to detect them.

In this Section, using all historic CME candidates we will examine the energy regime between the solar events and the Algol extreme flare and associated CME by populating the CME-flare diagram presented in Figure 5 of \citet{Moschou:17}. For that, we need a robust unified method to analyze and incorporate all the historic events so far in that diagram. In short, we use empirical relations to convert optical and UV fluences into X-ray flaring fluences and plot the Doppler-shift CME candidates in the same plot as the X-ray events.  We try to keep our assumptions to a minimum and always in the ranges provided in literature, when we calculate any quantities, such as CME mass and kinetic energy, that are not provided in the original papers. 
A summary of all the analyzable events examined in the current study can be found in Table \ref{tab:cme}.

\subsection{Events observed through Doppler-shifts}
\label{subsec:convertion}

In a statistical analysis \citet{Kretzschmar:11} concluded that for the Sun the flaring energy on the GOES soft X-ray passband constitutes about 1\% of the total radiated energy. Furthermore, emission in wavelengths below 50 nm make up about 10\% to 20\% of the that energy. While finally, the total visible and near UV parts of the spectrum contain most of the flare energy. It is not straightforward to conclude whether this relation extends to other stars.

Large statistical studies in both Sun-like and more active stars \citep{Butler.etal:88,Butler:93,Martinez-Arnaiz.etal:11} derived empirical relations between X-ray luminosity $L_X$ and the luminosities of optical lines, such as $H_\alpha$ and $H_\gamma$ lines, i.e. $L_{H\alpha}$ and $L_{H\gamma}$, respectively. More specifically, Balmer lines can be converted using the linear relation 
\begin{equation}
\label{eq:empirical}
L_X=32L_{H\gamma}=16L_{H\alpha}
\end{equation}
obtained/verified by multiple space and ground based telescopes to soft X-rays integrate in the range (0.04-2.0 keV), as described in \citet{Butler.etal:88} and further extended to a larger range in \citet{Butler:93}. Later on \citet{Martinez-Arnaiz.etal:11} examined a sample of about 300 late-type single stars with spectral types FGKM accounting for the first time for basal chromospheric contributions. From their full sample they found 243 counterparts with X-ray observations. They showed that there is no universal flux - flux relation for the chromospheric and coronal fluxes, with dK, dKe, and dMe\footnote{The symbols d before the M and K stars indicate a subdivision of those late-type dwarf star types, with dMe, dKe being stars with $H_\alpha$ in emission, while dK and dM being stars with $H_\alpha$ in absorption.} stars deviating from the general trend followed by less active stars. More specifically, \citet{Martinez-Arnaiz.etal:11} arrived to the relationship
\begin{equation}
\label{eq:fluxes}
\log{f_X}=(-2.19\pm0.41)+(1.48\pm0.07) \log{f_{H_\alpha}}
\end{equation}
associates $H_\alpha$ ($f_{H_\alpha}$) and X-ray ($f_X$) fluxes by fitting the data to a power-law relation \citep[see, Fig. 5 in][]{Martinez-Arnaiz.etal:11}, using the linear regression presented in \citet{Isobe.etal:90}. In other words, Figure 5 in \citet{Martinez-Arnaiz.etal:11} indicates that in the low energy regime $F_{H_\alpha}$ and $F_X$ are of the same order of magnitude, but this relation changes as we approach the X-ray saturation regime with the X-ray flux becoming one order of magnitude larger than the $H_\alpha$ flux. The \citet{Martinez-Arnaiz.etal:11} relation (Equation (\ref{eq:fluxes})) is more elaborate than the \citet[][]{Butler:93} one (Equation (\ref{eq:empirical})). However, our data set is so inhomogeneous that to avoid further discrepancies we use the \cite{Butler:93} relation (Equation (\ref{eq:empirical})) to convert the CME candidates that were observed with the Doppler-shift method into X-rays fluxes.

 In some of the cases discussed in this paper the observational passbands are in different wavelengths in the optical/UV regime instead of the $H_\alpha$ or $H_\gamma$ fluxes, which have well-known relations with X-ray emission. The \citet{Butler:93} relation indicates that the X-ray luminosities are typically of the order of a few tens of the optical luminosities observed. For those cases we multiply the observed optical/UV fluxes by 10 to estimate the X-ray flux inspired by Equation (\ref{eq:empirical}) and the high energy limit of Equation (\ref{eq:fluxes}). For one case \citep{Katsova.etal:99} only the broadband luminosity is known. In that case, we divide the broadband luminosity by a factor 10 inspired by the conclusions of the solar statistical study on the flare energy partition presented in \citet[][]{Kretzschmar:11}.


Recently, \citet{Vida.etal:19} performed a statistical analysis using archival stellar spectra and looked for line asymmetries. They found about 500 events with such asymmetries. As they note most of those events showed enhanced Balmer-line asymmetries, pointing to an association with flares. \citet{Vida.etal:16} observed speeds of the order of 100 \,--\, 300 km s$^{-1}$ and masses in the order of $10^{15}-10^{18}$~g. The authors were then able to fit their results with a log - linear relation between event masses and speeds that writes: 
\begin{equation}
\label{eq:vida}
log M_{blue}=(12.67\pm0.17) v_{blue}^{0.050\pm0.003} {,}
\end{equation}
with $M_{blue}$ being the mass of the blue-shifted material and $v_{blue}$ its blue-shifted speed. Equation~(\ref{eq:vida}) indicates that the masses and speeds of more ''energetic'' events increase simultaneously with masses increasing much faster and extending to more than 3 orders of magnitude than event speeds which only differ by a factor of a few. Individual event characteristics measurements were not reported in \citet{Vida.etal:19} and not all the events analyzed therein were associated with Balmer-line enhancements. Thus, we cannot include them in the current analysis.

\subsection{Events observed through X-ray absorption }
For the CME candidates observed through X-ray continuous absorption we follow a similar approach as the one presented in \citet{Moschou:17}. More specifically, we use the CME cone model, which is a geometric model often used for the analysis of CME events in the solar context \citep[see, e.g.][]{Howard82,Fisher1984,Zhao2002,Xie2004} to estimate the CME characteristics. The CME cone model requires a density for the transient plasma, a characteristic length and time scale, to determine the CME cruising speed, an opening angle of the CME cone and a thickness for the front of the conical shell. For the estimation of the CME mass we use the column density increase captured in each observation. As the characteristic time we choose the half-time of the column density decay i.e. $\tau = t(N_{H,max}/2)$, when possible. 
Then, based on the particular event we define a lower (flaring/obscuring) length scale to be equal to the flaring loop size and an upper (dynamic) length scale for the CME to be equal to 5 times the stellar radius \citet[se, e.g.,][]{Moschou:17}. 
In terms of opening angles we assume a semi-opening angle of $90^\circ$, since as was argued in \citet[][]{Aarnio2011SoPh} energetic events have wider opening angles and as was shown in \citet[][]{Moschou:17} there is a less than a factor 2 difference between our calculations for $75^\circ$ and $90^\circ$. Finally, we define the conical shell's thickness as 1/5 of each length scale, i.e. $L_{obs}/5$ and $L_{dyn}/5$. 
All the equations used for our analysis are detailed in \citet[][]{Moschou:17}.

\section{Stellar CME Candidates}
\label{sec:events}




\subsection{Doppler shift method} \label{ss:doppler}

CME candidates detected with the method of Doppler-shifted emission, mainly in Balmer lines, are presented in Table~\ref{tab:doppler}. The observed temperatures indicate that it is chromospheric material that is moving away from the host star. Here we describe the main characteristics of each event currently available. For the evaluation of the possibility of a Doppler based method we are interested in the mass and outflow speed measured at a particular astrocentric height of the host star. We can then estimate the local star-specific escape speed at the height of the CME measurement and compare it with the observed CME speed in order to gain insight on how likely that particular event is to escape the gravitational attraction of the star \citep[see e.g.][]{Guenther.Emerson:97,Bond.etal:01,Fuhrmeister.Schmitt:04}. The events are mentioned in chronological order of their occurrence. It is important to note that Doppler shift measurements only serve as \textit{lower limits} in the observed outflow speeds as they only measure the velocity component along the line of sight, thus suffering from large errors due to projection effects (for more see Section~\ref{sec:discussion}). It is difficult to follow an escaping CME in chromospheric lines as its mass will eventually get heated as it mixes with the hot $\geq 10^6 \mathrm{K}$ stellar corona. In Subsection \ref{sec:analysis} we discuss in detail how we analyze these events and convert their fluence to X-ray fluence. All analyzable events are also included in Figure \ref{fig:solar} with the non-escaping ones as lower limits and different colors.

\begin{table*}
\setlength{\tabcolsep}{4pt}
 \begin{center}
  \caption{Table with summary of all events potentially linked to CMEs observed through Doppler-shifts.} 
  \label{tab:doppler}
\begin{tabular}{|cccccccccc|}
 \hline
Refs &  Star & Type & D (pc) & Instrument & $u_{esc}$  & $u_{blue}$ & M $(\mathrm{g})$ & $F$ (erg) & $E_k$ (erg) \\
 \hline
1 & AD Leo & M4Vae & 5 & ESO, IDS & 580 & 1500-5830 & $>7.7\times10^{17}$ & $5\times10^{31}$ & $9\times10^{33}$ $^\mathrm{a}$ $-10^{35}$ \\
2 & AT Mic & M4.5Ve + M4.5Ve & 11 & \textit{SAAO}& 500 & 600 & $10^{14}-10^{16}$ &  $3\times10^{31}$ & $2-200\times10^{29}$ \\
3 & Cham J1149.8-7850 & wTTs & 140$^\mathrm{b}$ & FLAIR II & 440 & 600 & $1.4-78\times10^{18}$ &  $2\times10^{33}-10^{34}$ & $2.5-140\times10^{33}$ \\
4 & AU Mic & M1VeBa1 & 10 & EUVE & 375 & 1400 & $10^{20}$ & $3\times 10^{35}$ & $10^{36}$  \\
5 & V471 Tau  & K2V+DA & 48 & GHRS & 550 & $>120$ & $>3\times10^{15}$ & -- & $>2 \times 10^{29}$ \\
6 & DENIS 1048-39   & M9$^\mathrm{c}$ & 4.6$^\mathrm{d}$ & DENIS & 550 & 100 & $3-30\times10^{16}$ & $9 \times 10^{29}$ & $1.5-15\times10^{30}$ \\
7 & AD Leo  & M4Vae & 5 & FUSE & 580  & 84 & $4-500\times10^{17}$ & $2\times 10^{31}$ & $1.5-150\times 10^{31}$ \\
8 & V374 Peg  & M3.5Ve & 9.1 & RCC & 580 & $675$ & $10^{16}-10^{17}$ & $\sim10^{33}$  & $2-20\times10^{31}$    \\

 \hline
 \end{tabular}%
  \end{center}
{Notes.} Column 1 indicates the original reference, column 2 the star observed, column 3 its spectral type, column 4 its distance in pc, column 5 the instrument, columns 6 and 7 escape speed and blue-shift speed in $\mathrm{km\ s^{-1}}$, column 8 the mass of the event, column 9 the observed total radiated energy, and column 10 the kinetic energy of the event $E_k$. All distances are calculated by the \textit{SIMBAD} database \citep[][]{Simbad:00} parallaxes unless otherwise indicated. Spectral types are also given by \textit{SIMBAD}. \\
      $^\mathrm{a}$ $E_k$ estimated in \citet{Houdebine.etal:90} \\
      $^\mathrm{b}$ Distance given in \citet{Guenther.Emerson:97}\\
      $^\mathrm{c}$ Spectral type classification as reported in \citet{Fuhrmeister.Schmitt:04} \\
      $^\mathrm{d}$ Distance given in \citet{Fuhrmeister.Schmitt:04} \\
      {References.} (1) \citet{Houdebine.etal:90}, (2) \citet{Gunn.etal:94}, (3) \citet{Guenther.Emerson:97}, (4) \citet{Katsova.etal:99}, (5) \citet{Bond.etal:01}, (6) \citet{Fuhrmeister.Schmitt:04}, (7) \citet{Leitzinger.etal:11}, and (8) \citet{Vida.etal:16}. 
\end{table*}

\subsubsection{AD Leo, March 1984}
The best representative case for observing outflows through Doppler shifts, and one of the most promising CME candidates so far, was captured by \cite{Houdebine.etal:90}. The observed star was AD~Leo with $M=0.42M_\odot$ and $R_\star=0.46R_\odot$, which is a very active M3.5V (dMe) dwarf that produced a very powerful flare. The observation is unique in capturing line of sight plasma speeds as high as $5800\ \mathrm{km\ s^{-1}}$.  

\citet{Houdebine.etal:90} captured the mass outflow signature as a large blue shift in the Balmer lines $H_\gamma$ and $H_\delta$, using the $3.6 \mathrm{m}$ telescope at the European Southern Observatory (ESO). The strong blue shift (outflow) was measured during the impulsive flare phase and lasted for a few minutes, while only a weak red wing enhancement was present indicating a simultaneous downflow.
Before the flare onset, a faint absorption signature was measured in Ca II H and K lines corresponding to a $100\ \mathrm{km\ s^{-1}}$ with respect to the quiescent emission signal.
The inferred opacity from $H_\gamma/H_\delta$ line ratio measurements suggests that the plasma in the flaring region was forced to expand rapidly wigh its initial larger-than-the-mean-chromospheric opacity gradually decreasing. The fast expanding plasma scenario is consistent with the large speeds observed. 

\citet{Houdebine.etal:90} emphasize that initially the $H_\gamma/H_\delta$ decrements of the flare and the flow are similar, which is evidence that the two plasma elements may come from the same atmospheric region. In the first few minutes after the flare onset there was a strong deceleration, with the speed transitioning from $5830\ \mathrm{km\ s^{-1}}$ to $3700\  \mathrm{km\ s^{-1}}$ and then it remained more or less constant. This could plausibly be the impulsive phase of a CME during which the CME speed is much larger than the stellar wind speed and it decelerates due to the drag force \citep{Vrsnak2004,Zic2015}. A deceleration due to the interaction with the stellar wind is consistent with the small gravitational effect that \citet{Houdebine.etal:90} estimated. The escape speed for AD Leo was calculated to be $u_{esc}\sim 580\ \mathrm{km\ s^{-1}}$. {A minimum speed component of about $\sim1500\  \mathrm{km\ s^{-1}}$ was also measured during the first few minutes.} 

\citet{Houdebine.etal:90} estimated a plasma density on the order of $n>10^{10} \mathrm{cm^{-3}}$ which led to an estimated CME mass of $M=7.7 \times 10^{17} \mathrm{g}$ and a kinetic energy of $E_k=5\times10^{34}$ erg, for a plasma with temperatures of $15,000-20,000$~K. From photometry \citet{Houdebine.etal:90} inferred a total radiated energy on the order of $F_{broadband,tot}=(3\times10^{28}-10^{35})$ erg. For a flux varying from $f_{H_\gamma}=6.23\times 10^{-13}~\mathrm{erg\ cm^{-2}\ s^{-1}}$ to $f_{H_\gamma}=1.7\times 10^{-13}~\mathrm{erg\  cm^{-2}\ s^{-1}}$  in 7~min for AD Leo we get an $H_\gamma$ fluence of $F_{H_\gamma}=5\times 10^{31}$~erg. Then using Equation (\ref{eq:empirical}) we get an average X-ray fluence of $F_X\approx2\times10^{33}$~erg. The CME scenario explored therein resulted in an event with mass and kinetic energy several times larger than any solar CME, which led the authors to name such an event a super-CME and conclude that in very active stars there is the possibility of such transient events.

{\cite{Leitzinger.etal:11} used archival and published data from the Far Ultraviolet Spectroscopic Explorer (FUSE) to study activity phenomena on the flare star AD~Leo in the FUV. They used observational work from \citet{Christian.etal:06}, where two different flares were captured. AD Leo produced a blue wing enhancement in the O~VI~1032~\AA\ line after the flare event. The analysis of \citet{Leitzinger.etal:11} favored the dynamic CME scenario to explain this emission feature. The authors were able to exclude other possibilities that could give rise to this feature, such as co-rotating gas clouds in a static scenario, as the enhancement was only captured in a single spectrum. A direct interstellar medium (ISM) column density observation by \citet{Wood.etal:05} was used to correct for (ISM) absorption. 
The upflow speed in the AD~Leo case was estimated to be $\sim84\ \mathrm{km\ s^{-1}}$, similar to the velocity of chromospheric evaporation in the solar transition region, which was measured to be $\sim100\ \mathrm{km\ s^{-1}}$ \citep{Teriaca.etal:03}. \citet{Leitzinger.etal:11} argue that in the case that projection effects are extreme, for example when the CME propagated at an angle of $90^\circ$ from the line of sight there will be no projected speed and for a measured speed of $\sim 100\ \mathrm{km\ s^{-1}}$ an actual CME speed of the order of $5730\ \mathrm{km\ s^{-1}}$ could be present. Finally, \citet{Leitzinger.etal:11} excluded a prominence scenario by calculating a corotation radius smaller than the estimated possible cloud height $R=27.9 R_* \approx 12 R_\odot$. The authors favored the scenario of a CME with extreme projection effects lowering its actual speed. Finally, if we assume a plasma density of $n_H=10^{10}-10^{11}\mathrm{cm^{-3}}$ and an emitting volume of $2.5\times10^{31}-3\times10^{33}\mathrm{cm^{-3}}$ the estimated mass is $M\sim4\times 10^{17}-5\times10^{19}$g. Also, from Fig.~1 in \citet{Leitzinger.etal:11} for a duration of 200s the AD Leo flare emitted an energy of $\sim2\times10^{31}$erg in the C III multiplet.
}

\subsubsection{AT Mic, May 1992}

A large flare with total emitted energy in the wavelength range 3,600--4,200 \AA\ of $3\times10^{31}\ \mathrm{erg}$ was observed in an active M dwarf star, AT Microscopii  (dM4.5e, $M_*=0.27M_\odot$, $R_*=0.41R_\odot$), by \citet{Gunn.etal:94}. Spectroscopic observations were performed using the \textit{RPCS} spectrograph of the \textit{SAAO} 1.9m telescope in May 1992. Both the flare and quiescent emission were clearly seen in $H_\delta$ and Ca II lines. A strong blue asymmetry, which is an indication of a bulk outflow, was captured in Balmer emission lines. \citet{Gunn.etal:94} note that $200-400\ \mathrm{km\ s^{-1}}$ upflows in Ca XIX and Fe XXV have been observed for solar events \citep{Antonucci.etal:82,Hara.etal:11}. A maximum line of sight speed of $\approx 600 \mathrm{km\ s^{-1}}$, which decreased as the flare gradually decayed, was estimated. The escape speed at the surface of AT~Mic is $500 \mathrm{km\ s^{-1}}$. Chromospheric electron densities were taken as $n_e\sim 10^{12}-10^{14}\ \mathrm{cm^{-3}}$ and the mass of the cool $T\sim 10^{4} \mathrm{K}$ plasma was estimated to be $10^{15} \mathrm{g}$, which then gave a kinetic energy of $E_k\sim 3\times 10^{29}\ \mathrm{erg}$. \citet{Gunn.etal:94} emphasized that the estimated plasma parameters have large uncertainties, however they concluded that the mass flow kinetic energy was significantly less than the emitted flare energy.

\subsubsection{wTTs Cham, J1149.8-7850}
\cite{Guenther.Emerson:97} observed simultaneously 18 classical and 18 weak line T~Tauri stars (wTTs) using the FLAIR~II spectrograph on the UK Schmidt telescope over 14 hours. The authors argued that wTTs generally show non-variable emission over timescales of $> 100$ hours. However, two flares were captured from wTTs, both of which showed a fast rise phase and a longer decay lasting about an hour in $H_\alpha$. 
The wTTs J1149.8-7850 showed a flare with a $24\ \mathrm{min}$ rise phase and a slower decline phase that lasted about $\sim 160\ \mathrm{min}$. A large blue asymmetry was measured with speeds reaching $-1000\ \mathrm{km\ s^{-1}}$ during the flare peak and a total energy reaching $\sim 2\pm0.7\times10^{33}$ erg in $H_\alpha$. The estimated temperature of the flaring material was 20,000~K to 30,000~K. 

The second flaring star was J1150.9-7411, whose flare had a $30\ \mathrm{min}$ rise time, with a decay observed over the following two hours. After this period, the emission had not yet reached quiescent levels and \citet{Guenther.Emerson:97} estimated that the decay phase went on for another two hours. Thus, only a lower limit to the total energy release of $\geq 6\pm2 \times 10^{32}$~erg was found. The plasma had a temperature $<$30,000 K and the blue and red wings remained unchanged during the event. The two flares from wTTs stars were 700 and 200 times more energetic than solar events that have not been observed to release more than $3\times10^{30}$~erg in $H_\alpha$ \citep{Somov:92}.

The J1149.8-7850 event had a pronounced blue asymmetry in its flare profile, with one component at the star rest speed and another at $-600\ \mathrm{km\ s^{-1}}$. This blue-shift was interpreted as a CME by \citet{Guenther.Emerson:97}, as the outflow was faster than the local escape speed, which for a wTTs is $v_{esc}=440\ \mathrm{km\ s^{-1}}$ for $M_\star=M_\odot$ and $R_\star=2R_\odot$.  \citet{Guenther.Emerson:97} estimated lower limits of the mass at $1.4-78\times10^{18}\ \mathrm{g}$, an emitting volume of $6.2\times10^{31}-3.5\times10^{33}\ \mathrm{cm^{-3}}$ and a kinetic energy of $2.5\times10^{33}-1.4\times10^{35}$~erg, using a density of $10^{10}\ \mathrm{cm^{-3}}$. The total flux in the optical regime was inferred to be $\sim 10^{34}$~erg. \citet{Guenther.Emerson:97} estimated a magnetic field of 10--200 G and they noted that the difference with respect to solar flares was the emitting volume and the energy release, and not the magnetic flux density.




\subsubsection{V471 Tau, October 1994}

The Goddard High Resolution Spectrograph (GHRS) on the Hubble Space Telescope (HST) was used by \cite{Bond.etal:01} to perform an ultraviolet spectroscopic observation of the precataclysmic binary V471~Tau, which consists of a white dwarf and a dK2 star and an orbital period of 12.51 hr  \citep{Guinan.Sion:84}. The 
K dwarf is tidally-locked with the WD, and consequently rotates 50 times faster than the Sun and is thus much more active. 

The white dwarf acts as a UV background in spectra of the system. 
Evidence for a transient were found in the form of \textit{absorption features} in the white dwarf emission. \citet{Bond.etal:01} argued that, since the V471 binary is a detached binary system in which neither star fills its Roche lobe, the CME evidence found must originate from processes similar to solar CMEs, or in other words similar to an event originating from a single star. Wide $Ly\ \alpha$ absorption wings were captured in the spectra.

\citet{Bond.etal:01} inferred a speed of $v>120\ \mathrm{km\ s^{-1}}$ for the absorber from the line of sight measured speed of $\approx20\ \mathrm{km\ s^{-1}}$, that was seen to cross almost transversely the line of sight of the observation. The duration of the obscuring mass flow was 1,600~s. \citet{Bond.etal:01} inferred an extent for the CME of the order of $\geq 1.9\times 10^{10}\ \mathrm{cm}$, i.e.\ $L_{CME}\sim0.28R_\odot\approx0.29R_\star$ and 
calculated possible launch paths for the CME. They found trajectories that leave the K2 dwarf at a specific orbital phase, pass in front of the white dwarf with a speed of $\sim 22\ \mathrm{km\ s^{-1}}$, triggering absorption features in the spectra and finally leaving the binary system. Other launch trajectories and speeds were also explored and \citet{Bond.etal:01} report optimal velocities in the range $450-500\ \mathrm{km\ s^{-1}}$, consistent with solar observations \citep{Yashiro2009}. \citet{Bond.etal:01} estimated the upper limit for the number density of the absorbing mass to be $n_H\approx6\times10^{11}\ \mathrm{cm^{-3}}$ and the mass to be $M_{CME}>3\times10^{15}$g.

\subsubsection{DENIS 1048-39, March 2002}
The DENIS~1048-39 M9 dwarf was discovered by the DEep Near Infrared Survey (DENIS) at a distance of $4.6\mathrm{
pc}$. DENIS~1048-39 has an age of 1-2~Gyr, a radius of $0.1R_\odot$, a mass of $0.09M_\odot$ and is at the lower mass limit of hydrogen burning objects. However, as its $H_\alpha$ variable emission shows, it still displays observable magnetic activity levels. 
\citet{Fuhrmeister.Schmitt:04} performed observations of DENIS~1048-39 using the Ultraviolet-Visual Echelle Spectrograph (UVES) of the ESO Kueyen telescope. A blue shift of the order of $100\ \mathrm{km\ s^{-1}}$ was observed in the $H_\alpha$ and $H_\beta$ lines. A blue shift asymmetry was observed in both flare and background spectra this being strong in the $H_\alpha$ line. 

The $H_\alpha$ emission corresponds to a  temperature of approximately $T=10^4$ K. The measured luminosity of $H_\alpha$ was estimated at $1.6\times10^{26}\ \mathrm{erg\ s^{-1}}$ for a 4.6~pc distance, which gives a fractional luminosity of $\log\left(L_{H_\alpha}/L_{bol}\right)=-4.0$. The observed half widths of the $H_\alpha$ and $H_\beta$ lines were always lower than $20\ \mathrm{km\ s^{-1}}$, which indicates that the emission comes from a finite region on the stellar surface and since for both lines there is evidence for the presence of two components, \citet{Fuhrmeister.Schmitt:04} concluded that the emission could originate from two active regions. 

The escape velocity of the M9 dwarf is $v_{esc}\sim 550\ \mathrm{km\ s^{-1}}$, while the observed projected velocity is $100\ \mathrm{km\ s^{-1}}$. However, the blueshift only provides a lower limit for the outflow.  After 20 min exposure \citet{Fuhrmeister.Schmitt:04} estimated a $1.7R_\star$ altitude for the mass, which is larger than the corotation distance, thus excluding a prominence scenario. \citet{Fuhrmeister.Schmitt:04} discuss the possibility that the measured projected velocity is a result of the integration through different velocities during a possible initial CME deceleration phase. For a Balmer line brightening lasting for 1.5 h we have a total radiated energy of $F_{H_\alpha}=9\times10^{29}$~erg. Finally, if we assume a Hydrogen plasma number density of $n_H=10^{10}-10^{11}\mathrm{cm^{-3}}$ the estimated CME mass is $M\sim3\times 10^{16}-3\times10^{17}$g and a kinetic energy of $1.5\times10^{30}-1.5\times10^{31}$~erg.

\subsubsection{V374 Peg, 2005}

V374 Peg is a 200 Myr old ultrafast rotating single M4 dwarf, with a rotation period of 0.44~days, and is located 8.9 pc away. 
\cite{Vida.etal:16} performed spectroscopic and photometric observations of V374 Peg using various observational facilities expanding over 16 years. 
They concluded that V374 Peg has a magnetic field and a starspot configuration that remain very stable over a timespan of 16 years. There was no indication of cyclic starspot activity.

On HJD 2453603 \citet{Vida.etal:16} observed three consecutive blue-wing enhancements in Balmer lines, which suggests that they are related and resemble sympathetic solar flares. The $H_\alpha$ line was observed to be constantly in intense emission. The three blue-wing enhancements occurred at a specific rotational phase, which indicates long-lived magnetic loop systems that trigger flares. \citet{Vida.etal:16} argued that all three events came from the same active region nest. The long duration of the observation ($>10\ \mathrm{hours}$) allowed for better understanding the intensity variability before and after the enhancements. 

 All three blue-wing asymmetries occurred during a single flare with flaring energy $\sim10^{33}$ erg and with the third event being the strongest, corresponding to projected speeds of $675\ \mathrm{km\ s^{-1}}$. For its estimated mass of $M=0.3M_\odot$ and radius of $R=0.34R_\odot$ \citet{Vida.etal:16} deduced an escape velocity for V374 Peg of $v_{esc}=580\ \mathrm{km\ s^{-1}}$. They argued that the two first precursors ($350\ \mathrm{km\ s^{-1}}$ each) could be failed eruptions, with the red wing enhancement indicating material falling back after the initial upflow. 
 
 {Failed eruptions are common phenomena in the Sun \citep[e.g.][]{Joshi.eta:13,Zuccarello.etal:17}. A strong asymmetry was observed in a failed flux-rope eruption by \citet{Joshi.eta:13}, with a CME core getting formed. While the CME managed to lift off it was then  observed to fall back towards the Sun. The scientists suggested that overlaying magnetic flux tubes covering part of the initial filament could be a reason for the ballistic behaviour of the CME. In recent work, \citet{Zuccarello.etal:17} studied the transition between eruptive flares (associated with a CME) and failed eruptions. They concluded that failed eruptions occur when the supporting field of the filament and the overlying field gradually reach a less anti-parallel relative direction due to continuous photospheric shearing motions.} 
 
 The last of the three V374 Peg events, however, that was inferred by \citet{Vida.etal:16} to have occurred almost along the line of sight, had a projected (but close to true) speed of $675\ \mathrm{km\ s^{-1}}$ and this material escapes the gravitational attraction of the star. The authors estimate a CME mass of $M_{CME}\sim 10^{16}-10^{17}$g (one order of magnitude accuracy as the authors noted) following the \citet{Houdebine.etal:90} analysis, assuming a temperature $T=2\times10^4\mathrm{K}$ and a density in the range $10^{10}-10^{12}\mathrm{cm^{-3}}$.
Then \citet{Vida.etal:16} adopted an X-ray luminosity of log $Lx\sim 28.4$ erg~s$^{-1}$ and they inferred the occurrence of 15--60 CMEs per day with masses $M_{CME}>10^{16}$~g. 


\vspace{\baselineskip}

\subsection{X-ray absorption method} \label{ss:xrays}

X-rays are a valuable activity indicator, with a well-observed trend of increasing stellar activity with faster rotation to activity levels up to $10^3$ times higher than the current solar one. This trend, however, reaches a saturation regime wherein faster rotators cannot exceed fractional X-ray luminosities of the order of $L_X/L_{bol}=10^{-3}$ \citep{Feigelson.etal:04,Schmitt.Liefke:04,Telleschi.etal:07,Wright.etal:11,Wright.etal:18} for both partially convective and M-dwarf stars. While stellar coronae are unresolvable with current X-ray telescopes, flaring loop characteristics can be inferred by analyzing the flare decay profiles \citep{Reale.etal:97,Reale.etal:07}.

Several X-ray events with continuous enhanced absorption have been observed in different late-type active flaring stars. The largest events have been observed on short period binary stars such Algol-like systems and RS CVn-type systems, as well as on young fast rotators \citep[e.g.][]{Graffagnino.etal:95,Grosso.etal:97,Tsuboi.etal:00}. The working hypothesis for these events is that a mass of plasma is obscuring the flaring region and while escaping the gravitational potential of the star it also expands causing a characteristic decay of the absorbing column density. 

In an earlier study, we analyzed the most prominent representative of such an event \citep{Moschou:17} by assuming that a CME was ejected. Here, we use the CME cone geometric model, similar to the analysis presented in \citet{Moschou:17} for the Algol CME, for all the X-ray absorption events observed so far (see Table \ref{tab:xabs}). 

\begin{table*}[htbp]
 \begin{center}
 \caption{A summary of all events potentially linked to CMEs observed through X-ray absorption.} 
 \label{tab:xabs}
 \begin{tabular}{|cccccccccc|}
 \hline
Refs & Star & Type & D (pc) & $R_\star$ ($R_\odot$) & Mission & $L_{peak}^{X-ray} (\mathrm{erg\ s^{-1}})$ & $F(\mathrm{erg})$ & $N_{H,max}(\mathrm{cm^{-2}})$ & T(K) \\
 \hline
9 & Algol B & K2 IV & 28 & 3.5$^\mathrm{a}$ & BeppoSAX &  & $10^{37}$ & $3\times 10^{21}$ & $1.4\times10^8$  \\
10 & Prox Cen & dM5.5Ve & 1.3 & 0.15$^\mathrm{b}$ & Einstein & $2\times 10^{28}$ & $3.5\times 10^{31}$ & $10^{20}$ & $2.7\times 10^7$  \\
 11 & Algol & B8 V + K2 IV & 28 & 2.9 + 3.5$^\mathrm{a}$ & ROSAT & $2\times 10^{32}$ & $7\times 10^{36}$ & $\sim 2\times 10^{19}$ &  $10^8$  \\
12 & V773 Tau & K3Ve & 128 & 4.17$^\mathrm{c}$ & ASCA & $10^{33}$ & $10^{37}$  & $4\times10^{22}$ & $10^8$  \\
13 & UX Arietis &  G5 IV + K0IV & 51 & 1.6 + 5.6$^\mathrm{d}$ &  BeppoSAX & $10^{32}$ & $\ge 5\times 10^{36}$ & $10^{20} $ & $10^8$  \\
14 & Pleiades & K3 V + K3$^\mathrm{e}$ & 120 & 0.7$^\mathrm{e}$ &  XMM-Newton & $\approx 10^{30}$   & $3\times 10^{34}$ & $> 10^{22}$ & $1.5\times 10^7$  \\
15 & $\sigma$ Gem & K1IIIe & 38 & 10.1$^\mathrm{f}$ &  XMM-Newton & $10^{32}$ & $4.24 \times 10^{37}$ & $2.7\times 10^{20}$ & $2\times10^8$  \\
 \hline
 \end{tabular}
 \end{center}
 {Notes.} Column 1 indicates the original reference, column 2 the star observed, column 3 its spectral type, column 4 its distance in pc, column 5 stellar radius, column 6 the X-ray observatory, column 7 the peak X-ray luminosity, column 8 the observed X-ray fluence, column 9 the inferred column density, and column 10 the temperatures which are based on plasma model fits performed in the original papers. 
  All distances are calculated by the \textit{SIMBAD} database \citep[][]{Simbad:00} parallaxes unless otherwise indicated. Spectral types are also given by \textit{SIMBAD}. \\
  $^\mathrm{a}$ Algol radius as reported in \citet{Richards:93}\\
  $^\mathrm{b}$ Proxima radius as reported in \citet{Kervella.etal:17}\\
  $^\mathrm{c}$ V773 Tau radius as reported in \citet{Tsuboi.etal:98}\\
  $^\mathrm{d}$ UX Arietis radius as reported in \citet{Hummel.etal:17}\\
  $^\mathrm{e}$ Pleiades spectral type and radius as reported in \citet{Briggs.Pye:03}\\
  $^\mathrm{f}$ $\sigma$ Gem radius as reported in \citet{Roettenbacher.etal:15}  \\
 {References.} Continuing enumeration from Table \ref{tab:doppler}. (9) \citet{Favata1999}, (10) \citet{Haisch.etal:83}, 
(11) \citet{Ottmann.Schmitt:96}, 
(12) \citet{Tsuboi.etal:98}, (13) \citet{Franciosini.etal:01}, 
 (14) \citet{Briggs.Pye:03}, and (15) \citet{Pandey.Singh:12}.
 \end{table*}

\subsubsection{Algol events}

\paragraph{August 1992 event}
{\cite{Ottmann.Schmitt:96} observed Algol for 2.4 binary orbits with \textit{ROSAT}/PSPC in the second half of August 1992.  A giant X-ray flare took place in the middle of the observation and lasted for half a binary orbit, i.e. about 1.5 days. To establish a baseline emission level and isolate the flare signal, \citet{Ottmann.Schmitt:96} adopted the quiescent light curve observed by \citet{Ottmann:94}. The beginning of the flare was considered t be $t_0=293,000$~s into the observation. The deduced flare rise phase lasted for 24,000 s, while the decay extended up to 100,000 s. The authors estimated an e-folding time from the light curve of $\tau=30,400$~s, i.e. more than 8 hours.

The \citet{Raymond.Smith:77} 1T thermal plasma model was applied to the data, indicating that the temperature and density peaked during the flare rise phase and reached $T\approx 10^8$ K and $n_e\approx 5\times 10^{11} \mathrm{cm^{-3}}$, respectively. The observations of \citet{Ottmann.Schmitt:96} are illustrated in the top middle panel of Figure \ref{fig:xray} with errors at 90\% confidence levels.

\citet{Ottmann.Schmitt:96} discussed a potential column density increase by a factor of 2 during the early decay phase, reaching $N_H\approx 2\times 10^{19} \mathrm{cm^{-2}}$. Similar to \citet{Haisch.etal:83}, the authors characterized the giant Algol flare as a two-ribbon one. The flare had a thermal energy of $7\times 10^{36}\mathrm{erg}$, a peak luminosity of $L_X=2\times 10^{32} \mathrm{erg\ s^{-1}}$ and a flaring volume of $V=1.3\times10^{34} \mathrm{cm^3}$. A flaring loop length of $5\times 10^{11} \mathrm{cm^{-3}}$, corresponding to an altitude of $0.65 R_B$, was estimated by applying a quasistatic cooling method. }

\paragraph{August 1997 event}
An X-ray stellar flare that occurred on Algol in 1997 August 30 ranks amongst the most energetic stellar flares ever observed. The proximal (28.5 pc) Algol prototypical system is a binary star with an early type primary which is a B8 V ($R_A=2.9 R_\odot$) star and a secondary K2 IV ($R_B=3.5 R_\odot$) star. The binary system is tidally locked and has a period of 2.7 days.

The flare observation was made by \textit{BeppoSAX}/LECS, MECS, PDS and the events was analyzed by \citet{Favata1999}, who found the total X-ray fluence to be approximately $1 \times10^{37}\;\mathrm{erg}$ in the 1--8~\AA\ GOES band. \citet{Favata1999} also measured a \textit{quiescent} volume emission measure of $EM=n_e^2V=3\times 10^{53} \mathrm{cm^{-3}}$.
Shortly after the Algol super-flare peak, the column density increased significantly (by $\sim$ 2 orders of magnitude) from the $\approx 10^{20} \mathrm{cm^{-2}}$ base value and then gradually decayed with an e-folding time of $\tau=5.6 \mathrm{ks}\approx 1.5\mathrm{hours}$ \citep{Moschou:17}. 

The scenario of a CME being responsible for the X-ray continuous absorption and the column density variability was explored by \citet{Moschou:17}. Those authors analyzed the column density profile and showed that it followed a $\propto t^{-2}$ law, which is consistent with a quasi-constant CME propagation and expansion. Using physical arguments, \citet{Moschou:17}, defined two length scales for the CME evolution, a dynamic one based on the length scale established by solar studies \citep{Zic2015}, where the CME pressure balances the solar wind one, and minimum length scale such that the CME just obscures the flaring region initially. Then, using a) the derived temporal relation for the column density ($N_H\propto t^{-2}$), which in that case was consistent with a CME expanding self-similarly, b) the geometric CME cone model often used in solar cases \citep{Howard82,Fisher1984,Zhao2002,Xie2004}, and c) geometric arguments about the flaring site, \citet{Moschou:17} obtained lower and upper limits for the CME mass and kinetic energy in the ranges $2\times 10^{21}$--$2\times 10^{22}$~g and $7\times 10^{35}$--$3\times 10^{38}$~erg, respectively.

\subsubsection{Proxima Centauri, August 1980}

A five hour observation of the dM5e star Proxima Centauri ($R_*=0.15 \mathrm{R_\odot}$) was performed by \citet{Haisch.etal:83} using simultaneously observations from \textit{Einstein}/IPC and the \textit{International Ultraviolet Explorer (IUE)} in August 1980. The authors report indirect evidence of a two-ribbon flare, which is considered an indication of a prominence eruption in solar physics. The quiescent coronal luminosity was measured to be $L_{cor}\sim 5\times 10^{26} \mathrm{erg\ s^{-1}}$ and was subtracted from the X-ray flare light curve. 
Prox Cen has an X-ray to bolometric luminosity ratio 100 times larger than the Sun, reaching $L_{cor}/L_{bol}\sim 8\times 10^{-5}-3\times 10^{-4}$.  
The coronal temperature was estimated to be  $T\sim 4\times 10^6$ K. For the flare a peak was measured at $L_x\approx 2\times 10^{28} \mathrm{erg\ s^{-1}}$ (consistent with an X20 solar flare according to the GOES\footnote{Solar flares are classified according to their energies into A, B, C, M and X-class events ranging from the weakest to the most energetic ones. The flare classification in the GOES scale is based on the flare peak emission in the soft X-ray band of 1 - 8 \AA.} (Geostationary Operational Environmental Satellite) scale), and a maximum temperature of $\sim 27\times 10^6$ K. The observation lasted for the entirety of the flare, capturing both the pre-flare and post-flare luminosity levels. A best fit $\chi^2$ thermal plasma model \citep[see, e.g.][]{Raymond.Smith:77} was used to calculate $T$ and $N_H$ at 90\% joint confidence levels, including all sources of uncertainty apart from the \textit{Einstein} instrument calibration errors. 

\begin{figure*}[htbp]
\begin{center}
\includegraphics[width=\textwidth]{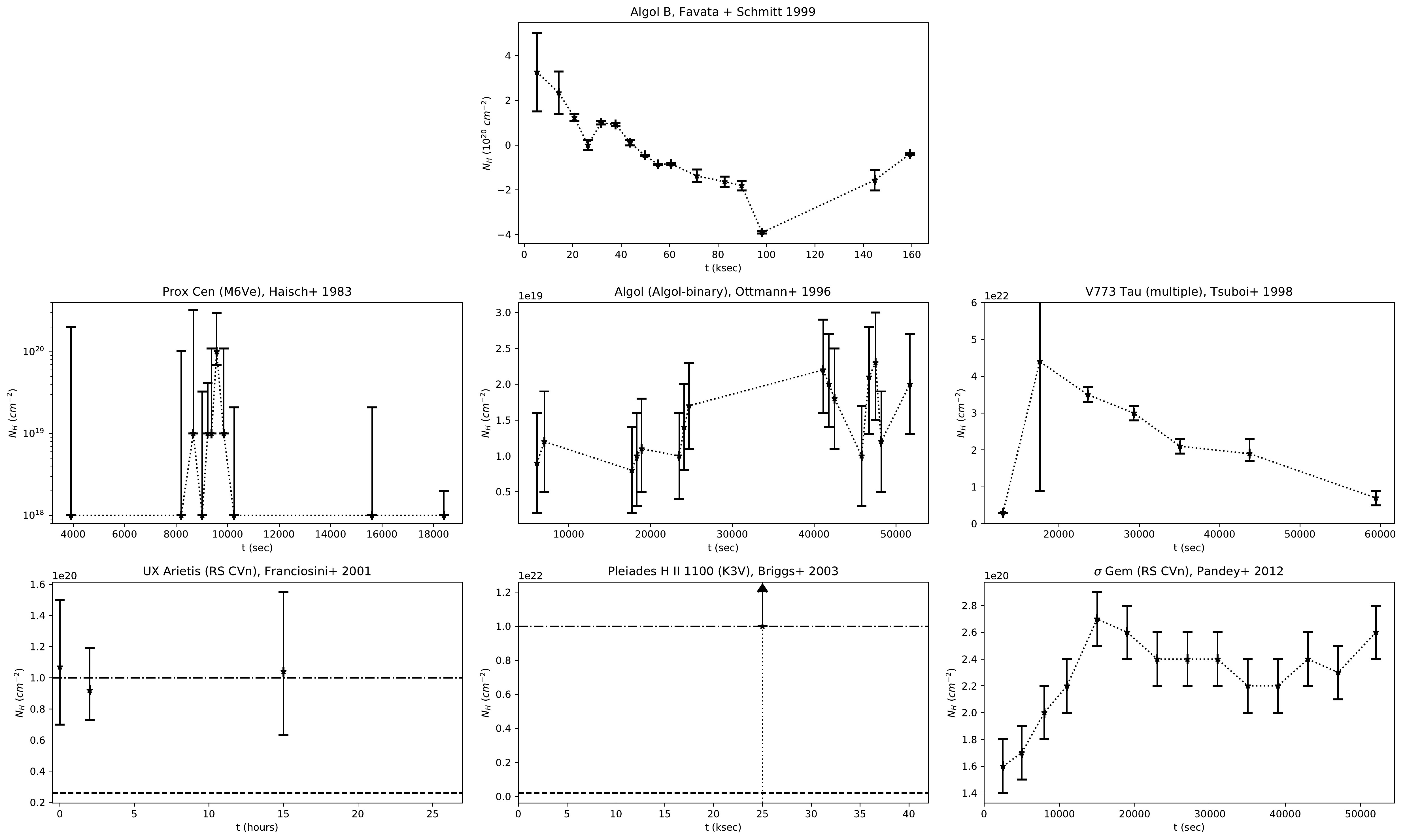}
\caption{All CME candidates associated with X-ray absorption including the Algol B event presented in \citet{Moschou:17}. The full references and details on each observation can be found in Table~\ref{tab:xabs}. We have off-set the observations in time to better capture the evolution of the column density $N_H$, i.e. the rise and decay phases.}
\label{fig:xray}
\end{center}
\end{figure*}

The estimated column density, $N_H$, with the errors reported in \citet{Haisch.etal:83}, is illustrated in the top left panel of Figure~\ref{fig:xray}. The column density increased immediately after the flare luminosity peak, reaching a value of $N_H\approx 10^{20} \mathrm{cm^{-2}}$ to then return to quiescent pre-flare values of the order of $10^{18}-10^{19} \mathrm{cm^{-2}}$. The authors report an e-folding time ($1/e$) for the temperature and luminosity of the order of $\tau \approx 20 ~\mathrm{min}$.

Using solar coronal loop models \citet{Haisch.etal:83} were able to estimate an X-ray emission measure of  $EM\approx 8.3\times 10^{48} \mathrm{cm^{-3}}$ for the coronal temperature, an upper limit for the flaring loop length $L\leq 2.4\times 10^9 \mathrm{cm}$, and a lower limit for the coronal density $n\geq 7.5\times10^9  \mathrm{cm^{-3}}$. By comparing the Prox Cen flare to solar two-ribbon flares \citet{Haisch.etal:83} were able to estimate the total flaring energy at $E_{tot}=3.5\times 10^{31}\mathrm{ergs}$. The authors discussed the possibility of cool, dense prominence material being the source of the sharp $N_H$ increase to $10^{20} \mathrm{cm^{-2}}$ shortly after the flare peak. This interpretation was consistent with solar two-ribbon flares that are often associated with transients and contain enough mass to cause the observed column density  increase. 

Finally,  using the simultaneous IUE ultraviolet measurements \citet{Haisch.etal:83} concluded that the radiative energy loss in the observed Prox Cen flare was more important in the corona rather than in the chromosphere or transition region, consistent with a dominant radiative cooling (rather than conduction cooling for the solar case) and maybe cooling due to expansion.


\subsubsection{V773 Tau, February 1995}

In February 1995 \cite{Tsuboi.etal:98} obtained a 40~ks ($\approx 11$ hours) observation of the weak-line T~Tauri star (wTTs) V773~Tau using \textit{ASCA}/SIS,GIS. V773~Tau is a spectroscopic binary consisting of K2~V and K5~V stars \citep{Welty:95}. A strong flare was observed in which the X-ray emission increased by a factor of 20. The flare had a sharp rise phase and a slower exponential decay phase with an e-folding time of $\tau=8.2  \mathrm{ks}\approx 2.3 \mathrm{hours}$. More specifically, \citet{Tsuboi.etal:98} reported a large \textit{lower limit} for  the peak flare luminosity of the order $L_X\approx 10^{33} \mathrm{erg\ s^{-1}}$ and a total energy reaching $E_{tot}\approx 10^{37} \mathrm{ergs}$, which were calculated after subtracting the quiescent X-ray spectrum obtained from their pre-flare measurements.

\citet{Tsuboi.etal:98} note that typical X-ray flares from TTs have luminosities $L_X\geq 10^{31} \mathrm{erg\ s^{-1}}$, i.e.\ 4 orders of magnitude more powerful than typical solar events \citep[e.g.][]{Montmerle.etal:83,Preibisch.etal:95,Kamata.etal:97}. Like solar flares, this rapid energy release is most likely due to magnetic reconnection. Strong magnetic activity in TTs is not only observed in X-rays, but also in radio wavelengths that reach luminosities of the order $L_R=10^{18} \mathrm{ergs\ s^{-1} Hz^{-1}}$ often accompanied by circular polarization during flare and quiescent times \citep[][]{Phillips.etal:91,White.etal:92,Skinner:93}. Circular polarization offers direct evidence for the presence of magnetic fields. 

After subtracting the preflare spectrum and using thermal plasma models, \citet{Tsuboi.etal:98} were able to obtain a best fit and estimate the temperature and column density at each time bin. We illustrate their results and corresponding errors in the top right panel of Figure~\ref{fig:xray}. The flaring material reached a maximum temperature of $T=10^8$ K, a maximum column density of $N_H=4\times10^{22}\mathrm{cm^{-2}}$, and a volume emission measure of the order of $EM\approx10^{55} \mathrm{cm^{-3}}$. The column density was significantly larger during the flare than in pre- or post- flare measurements. An agreement of broadband X-ray fluxes between \citet{Tsuboi.etal:98} and \citet{Skinner:93} indicate that V773~Tau has a constant X-ray emission level with $L_X/L_{bol}=10^{-3}$.  \citet{Tsuboi.etal:98} noted that the extreme energy release associated with the flare could cause expansion or evaporation of underlying material. Then, a quasistatic cooling loop model gave an electron density of  $n_e=3\times10^{11}\mathrm{cm^{-3}}$, a derived emitting volume of $V=6\times10^{32}\mathrm{cm^{3}}$ and a loop size of $L=4\times 10^{11} \mathrm{cm}$, which is larger than the stellar radius $R_\star=2.9\times 10^{11} \mathrm{cm}=$, but still 10 times smaller than the semi-major axis of the system.

\subsubsection{UX Arietis}

UX Ari consists of a G5 IV and a K0 IV star and is one of the most active RS~CVn systems known.
Observations in two sequential years, August 1997 and August 1998, of UX Ari with \textit{BeppoSAX} and all three LECS, MECS and PDS detectors were presented by \cite{Franciosini.etal:01}. 
An hour-long flare was observed during the first observation, but just a quiescent signal was picked up during the second one.
Only the final 2 hours of the flare rise phase were captured, as the flare appeared to have started before the observation began. The flare decay phase was longer, with an e-folding time $\tau\approx 13 \mathrm{hours}\approx 47 \mathrm{ks}$. \citet{Franciosini.etal:01} note that a contribution from rotational modulation cannot be ruled out, since the observations cover only slightly less than half an orbital period (0.4). 

The spectra were fitted with a two temperature thermal plasma models using \textit{MEKAL} \citep{Mewe.etal:95} and accounting for interstellar absorption. The best fit gave a column density of $N_H=2.6\times10^{19}\mathrm{cm^{-2}}$ for the quiescent (second) observation of August 1998 and $N_H\approx\times10^{20}\mathrm{cm^{-2}}$ for the entire flare duration during August 1997. In Figure~\ref{fig:xray} we illustrate at the bottom left panel the column density profile from the best fit \citet{Franciosini.etal:01} and the errors reported therein. \citet{Franciosini.etal:01} used the updated response matrix released in January 2000 for the \textit{LECS} instrument which solved the problem of systematically overestimating hydrogen column densities when observing stellar coronae with BeppoSAX. With the matrix giving a quiescent column density of $N_H=8\times10^{19}\mathrm{cm^{-2}}$, i.e. 3 times larger than the updated matrix, similar to what was found for all previous \textit{BeppoSAX} observations including \citet{Favata1999}. However, the flare column density even through it is 1.4 times lower than the one obtained with the old matrices it is still pretty high and more specifically $\sim 5$ times larger than the quiescent value. 

The hard X-ray emission is fully attributed to the flare related hot plasma. The peak X-ray luminosity was $L_X\approx 10^{32} \mathrm{erg\ s^{-1}}$, reaching a value of $1.4\times 10^{31} \mathrm{erg\ s^{-1}}$ at the end of the flare.  The peak emission measure was $EM\sim 8\times 10^{54} \mathrm{cm^{-3}}$ and the peak temperature $T=100$~MK. The total X-ray energy release during the flare was $E_X\geq5\times 10^{36} \mathrm{erg}$.

A two ribbon flare model \citep{Poletto.etal:88} adopted from the solar original \citep{Kopp.Poletto:84} was used to analyse the data, in which the magnetic energy released depends on the flaring region and the magnetic field strength. Constraining the surface magnetic field strength to of kG order, according to other observations in RS CVn-type binaries, \citet{Franciosini.etal:01} discuss that the flare took place in a $33^\circ - 53^\circ$ region corresponding to a $\sim 0.6 - 0.9 R_\star$ width. The flaring loop half length was estimated at  $\sim 0.1 - 0.2 R_\star$ during the flare peak and at  $\sim 0.5 - 0.8 R_\star$ at the end of the flare. These results were in agreement with the ASCA flare analyzed by \citet{Gudel.etal:99}. For these estimations the authors assumed an evaporation velocity of $500 \mathrm{km\ s^{-1}}$. The X-ray emission amounted to only a small 0.1$-$0.3\% fraction of the total magnetic energy release. Finally, \citet{Franciosini.etal:01} theorize that a CME and resulting absorbing material along the line of sight could explain the column density increase during the flare by a factor 5 with respect to the quiescent observation.

\subsubsection{LQ Hyades, November 1992}
\cite{Covino.etal:01} used \textit{ROSAT}/PSPC and \textit{ASCA}/SIS,GIS to observe GI 355, an active young star in LQ Hya in X-rays. The \textit{ROSAT} observations were divided in eight parts lasting for 1,500 to 2,000 s each. \textit{ROSAT} observed a large flare with a peak X-ray flux of $10^{31}\mathrm{erg\ s^{-1}}$. The decay phase of the flare had an e-folding time $\tau\sim 10.1 \mathrm{ks}$ \citep{Covino.etal:01}. Roughly six months after the \textit{ROSAT} observations, \textit{ASCA} observed GI 355 for about 20,000 s. For each \textit{ROSAT} pointing, 1 and 2 temperature models with free metallicity $Z$ and column density $N_H$  were used to fit the data. \citet{Covino.etal:01} note that they were not able to get any satisfactory spectral fit during the main part of the flare for \textit{ROSAT}. Fits with 1, 2 and 3 temperature models, fixed column density $N_H=4\times10^{19}\mathrm{cm^{-2}}$ and free metallicity were performed for \textit{ASCA} data. The column density remained constant during the flare without any increase from flare onset to flare decay. The total emitted X-ray energy was $F_X=9\times10^{34} \mathrm{erg}$. \cite{Covino.etal:01} used the approximate density for the interstellar material found in \citet{Paresce:84} $n_H=0.07 \mathrm{cm^{-3}}$ for obtaining an estimate of the column density of a star at 18 pc (like GI 355) and got a column density of $N_H\sim 4\times 10^{18} \mathrm{cm^{-2}}$, i.e. one order of magnitude lower than the fitted value. However, \citet{Covino.etal:01} note that they found no significant variation in the column density temporal profile to indicate the presence of a CME, but only an increased value with respect to the value from the \citet{Paresce:84} paper. We cannot analyze this event as a CME based on this analysis, as even if the column density is increased there is no indication of a mass outflow. From solar physics we know that CMEs do not always erupt during the flare onset at the flare site, however an outgoing flow would present some decay with time if it were indeed captured in the flare spectra. However, the presence of prominence material cannot be excluded.


\subsubsection{Pleiades, H II 1100, September 2000}

The Pleiades is a nearby young cluster with an age of $\approx 100\mathrm{Myr}$, consisting of stars ranging from B to M types.  These characteristics make it a useful target for studying X-ray producing processes in coeval stars with different masses.  Indeed, 
several X-ray telescopes have targeted Pleiades including \textit{Einstein}, \textit{ROSAT}, and \textit{Chandra} \citep{Micela.etal:90,Stelzer.Neuhauser:01,Daniel.etal:02}.

Of more interest for our study, the Pleiades was also observed in a 40 ks X-ray survey with  \textit{XMM-Newton}/EPIC by \cite{Briggs.Pye:03}. 
 During the observation, the H II 1100 binary star (\citet{Briggs.Pye:03} used H II number designations from \citet{Hertzsprung:47}, Table 2) consisting of twin K3 stars produced a flare with decay time $\tau\sim 3\ \mathrm{ks}$. The fractional quasi-steady X-ray luminosity of H II 1100 was measured to be $L_X/L_{bol}=-3.48$. The derived plasma temperature was in the range 14--17~MK with an average of 15.4 MK, while the emission measure and X-ray flux at the flare peak were estimated to be $EM=1.4\times10^{53}\ \ cm^{-3}$ and $L_X\approx10^{30}\ \mathrm{erg\ s^{-1}}$, respectively. The flaring loop size was then inferred to be $1.2\times10^{10}\ \mathrm{cm}\sim 0.25R_\star$ using a hydrodynamic flare model scaling technique similar to \citet{Reale.etal:97}. A mean electron density in the flaring loop at flare peak was estimated at $n_e\sim 5\times10^{11}\ \mathrm{cm^{-3}}$, while the required magnetic field strength for confinement was $B\sim 300\ \mathrm{G}$.

A sharp dip in the H II 1100 light curve was extensively discussed in \citet{Briggs.Pye:03}, who examined various explanations including an eclipse from an orbiting Jupiter size planet, obscuration of the flaring site by a CME, and the coincidence of a two flares with the first having a faster decay than its rise. \citet{Briggs.Pye:03} did not favor either of these over the others. They noted that an eclipse by a Jupiter like body eclipse has a low probability of occurrence. For a CME explanation, 
a column density of the order of $N_H>10^{22}\ \mathrm{cm^{-2}}$, which is two orders of magnitude higher than a typical solar prominence value, and velocities of $\sim150-500\ \mathrm{km\ s^{-1}}$, are required. Figure~\ref{fig:xray} (bottom right panel) shows the nominal column density for Pleiades and the required one for a CME obscuration event. \citet{Briggs.Pye:03} estimate that such a CME would have a kinetic energy of the order of $E_k\sim 6\times10^{33}\ \mathrm{erg}$, which is comparable to the total flaring energy observed. Finally, assuming a flare duration of about 30~ks and a constant flux of $L_X\approx10^{30}\ \mathrm{erg\ s^{-1}}$, the total flaring energy becomes $E_{tot}=3\times10^{34}$~erg.

\subsubsection{$\sigma$ Gem, April 2001}
Seven flares from five binary systems of RS CVn type were observed by \citet{Pandey.Singh:12}  using \textit{XMM-Newton} and the European Photon Imaging Camera (EPIC). with the strongest flare observed in the RS CVn $\sigma$ Gem.   $\sigma$ Gem has an active K1 III type primary and an unobserved late-type main sequence secondary.  $\sigma$ Gem was observed for 56 ks. During these measurements only the flaring without the quiescent states were observed for  $\sigma$ Gem. A quiescent state was then estimated using the \citet{Nordon.etal:06}  observation of $\sigma$ Gem with \textit{Chandra}. \citet{Pandey.Singh:12}, thus adopted a value of 25\% lower flux with respect to the measured flare peak for the quiescent emission of $\sigma$ Gem. The X-ray peak luminosity for the $\sigma$ Gem flare is $L_X \approx 10^{32} \mathrm{erg\ s^{-1}}$, with a total emitted energy of the order of $F_X=4.24\times 10^{37} \mathrm{erg}$, and an e-folding flare decay time of $\tau\approx32.5 \mathrm{ks}$. For obtaining the best fit in the $\sigma$ Gem data all one (1T), two (2T) and three (3T) temperature collisional plasma models (\textit{APEC}) were used \citep{Smith.etal:01}. However, 1T and 2T models using solar photospheric abundances could not fit the data well, thus a best-fitting 3T model was used for $\sigma$ Gem. The derived column density was found to increase by a factor $\sim 2$ during the flare rise phase and decrease during the flare decay phase. The maximum column density reached $N_H=2.7\times 10^{20} \mathrm{cm^{-2}}$, i.e. twice as high as the quiescent value. Figure~\ref{fig:xray} bottom middle panel shows the column density profile estimated in \citet{Pandey.Singh:12}. The temperature and metallicity appear to  peak at the flare onset phase. The emission measure reached a value of $EM=3.77\times 10^{54} \mathrm{cm^{-3}}$, i.e. increased by a factor $\sim 1.8$. A time-dependent hydrodynamic model \citep{Reale.etal:97,Reale.etal:04,Reale.etal:07} was employed similar to \citet{Pandey.Singh:08} to estimate characteristics of a single flaring loop. The peak temperature for the $\sigma$ Gem flare was $177\pm6 \mathrm{MK}$. The hydrodynamic flaring loop length was estimated to be $<6.43\times10^{11} \mathrm{cm}$, while the rise and decay based estimations gave $>2.55\times10^{11} \mathrm{cm}$ and $5.18\times 10^{11} \mathrm{cm}$, respectively. Assuming a fully ionized hydrogen plasma an electron density of the order $10^{10-11}\mathrm{cm^{-3}}$ was found.

\begin{table*}[htbp]
 \begin{center}
  \caption{Analyzable stellar CME candidates using Doppler-shift (top half) and X-ray (bottom half) methods.}
 \label{tab:cme}
 \begin{tabular}{|c|c||ccc|cc|c|}
 \hline

Blueshift & Star &  $u_{CME} \mathrm{(km~s^{-1})}$ & $M_{CME}$ (g) & $E_{k,CME}$ (erg) & Emission & F (erg) & $F_X$ (erg)  \\
 \hline
1 & AD Leo & 1500-5830 & $>7.7\times 10^{17}$ & $9\times 10^{33}-10^{35}$ & $H_\gamma$ & $5\times10^{31}$  & $8\times10^{32}-3\times10^{33}$  \\
2 & AT Mic & 600 & $10^{14}-10^{16}$ & $2-200\times10^{29}$ & $H_\delta$ & $3\times10^{31}$ & $3\times10^{32}$ \\
3 & wTTs Cham  & 600 & $1.4-78\times10^{18}$ & $2.5-140\times 10^{33}$ & $H_\alpha$  & $2\times10^{33}-10^{34}$ & $3.2\times10^{34}-1.6\times10^{35}$   \\
4 & AU Mic & 1400 & $10^{20}$ & $10^{36}$  & BB$^\mathrm{a}$ & $3\times 10^{35}$ & $\sim3\times10^{34}$ \\
6 & DENIS 1048-39 & 100 & $3-30\times10^{16}$ & $1.5-15\times10^{30}$ & $H_\alpha$, $H_\beta$ & $9 \times 10^{29}$ & $9 \times 10^{30}$ \\
7 & AD Leo & 84 & $4-500\times10^{17}$ &  $1.5-150\times10^{31}$  & C III & $2\times 10^{31}$ & $2\times10^{32}$ \\
8 & V374 Peg  & $675$ & $10^{16}-10^{17}$ & $2-20\times10^{31}$ & $H_\alpha$ & $\sim10^{33}$  &  $\sim1.6\times10^{34}$ \\
 \hline
 \hline
X-rays  & Star &  $u_{CME} \mathrm{(km~s^{-1})}$ & $M_{CME}$ (g) & $E_{k,CME}$ (erg) & $\tau_{1/2}$ (ks) & L ($R_\odot$) & $F_X$ (erg) \\
 \hline
 9 & Algol B$^\mathrm{b}$  & 250-6600 & $2\times10^{21}-2\times10^{22}$ & $7\times10^{35}-3\times10^{38}$ & {5.6} &  & $10^{37}$ \\
10 & Prox Cent  & 40-1000 & $1.2\times10^{16}-7\times10^{18}$ & $ 10^{29}-4\times10^{34}$ & 0.5& 0.03-0.75 & $1.7\times 10^{31}$ \\
11 & Algol  & 280-2400 & $6\times10^{19}-4.5\times10^{21}$ & $2.3\times10^{34}-1.3\times10^{38}$ & 5 & 2-17.5 & $1.15\times 10^{37}$ \\
12 & V773 Tau  & 210-730 & $10^{20}-1.4\times10^{21}$ & $2.5\times10^{34}-3.7\times10^{36}$ & 20 & 6-21 &  $8.8\times10^{36}$ \\
13 & UX Arietis &    & prominence?  & &   &   &  $\ge 3.7\times 10^{36}$ \\
14 & Pleiades &    &  prominence?  &   &   &  & $8.3\times 10^{33}$ \\
15 & $\sigma$ Gem  & 500-2800 & $3\times10^{21}-10^{23}$ & $4\times10^{36}-4\times10^{39}$ &  12.5 & 9-51 & $6.4 \times 10^{37}$ \\
 \hline
 \end{tabular}
 \end{center}
  {Notes.} The first 5 columns are the same both observational methods. Column 1 indicates the original reference, column 2 the star observed, column 3 the derived CME speed, column 4 the derived CME mass, column 5 the CME kinetic energy. \textit{Doppler-shifts}: Column 6 indicates the emission passband, column 7 the total measured fluence, and column 8 the converted X-ray fluence. \textit{X-ray absorption}: Column 6 indicates the time for the column density to reach half its maximum value, column 7 the flaring loop length, and column 8 the X-ray fluences converted to the GOES passband (1\,--\,8 \AA) using the Chandra/PIMMS online tool and setting the plasma model to MEKAL, the galactic column density to $N_H=10^{20}\mathrm{cm^{-2}}$ and the abundance to 0.4 times the solar value. For more details please advise the manuscript. \\
    $^\mathrm{a}$ Broadband\\
      $^\mathrm{b}$ CME characteristics taken from \citet{Moschou:17} \\
      {References.} As presented in Tables \ref{tab:doppler} and \ref{tab:xabs}.
 \end{table*}

\begin{figure}[htbp]
\begin{center}
\includegraphics[width=\columnwidth]{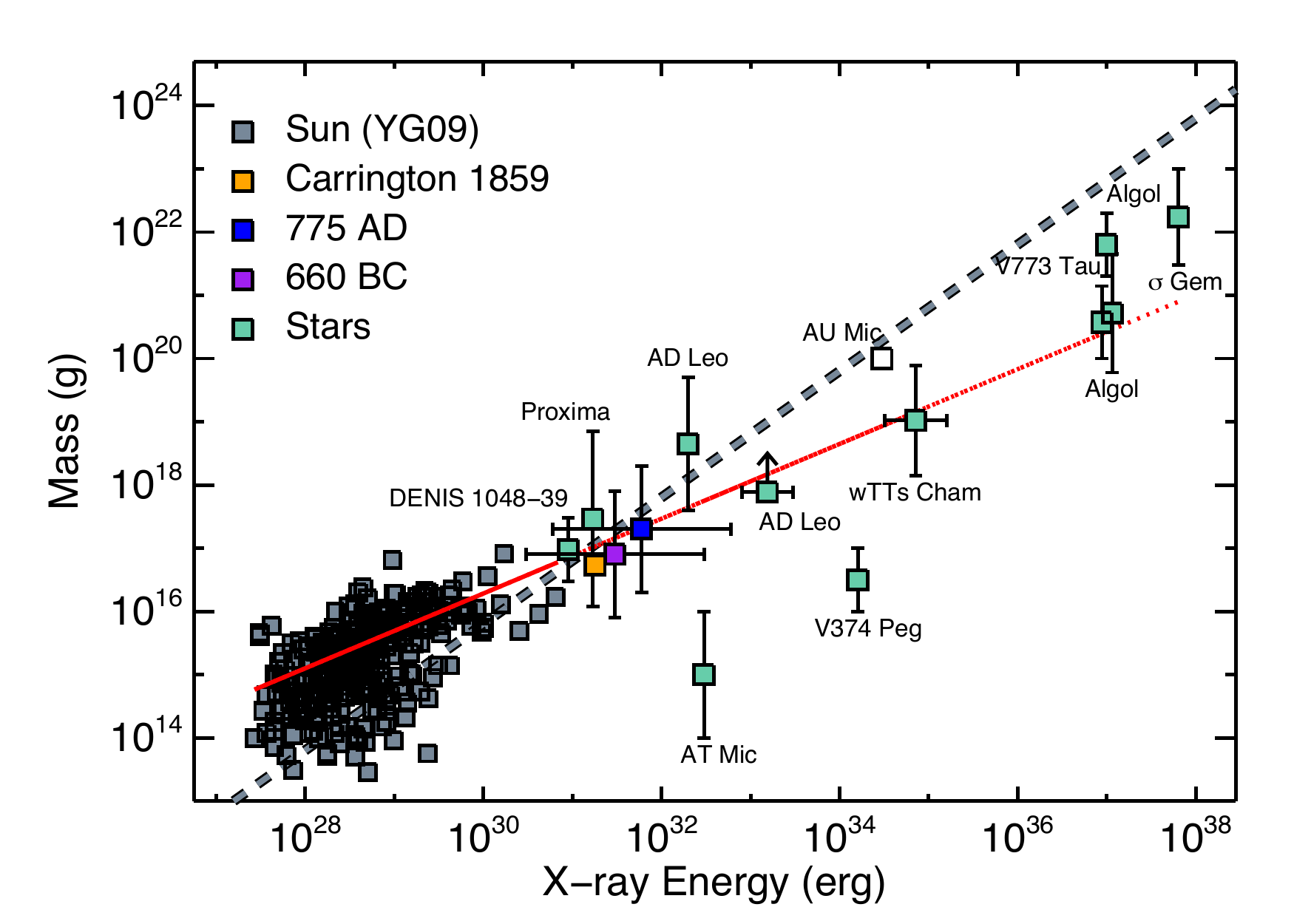}
\includegraphics[width=\columnwidth]{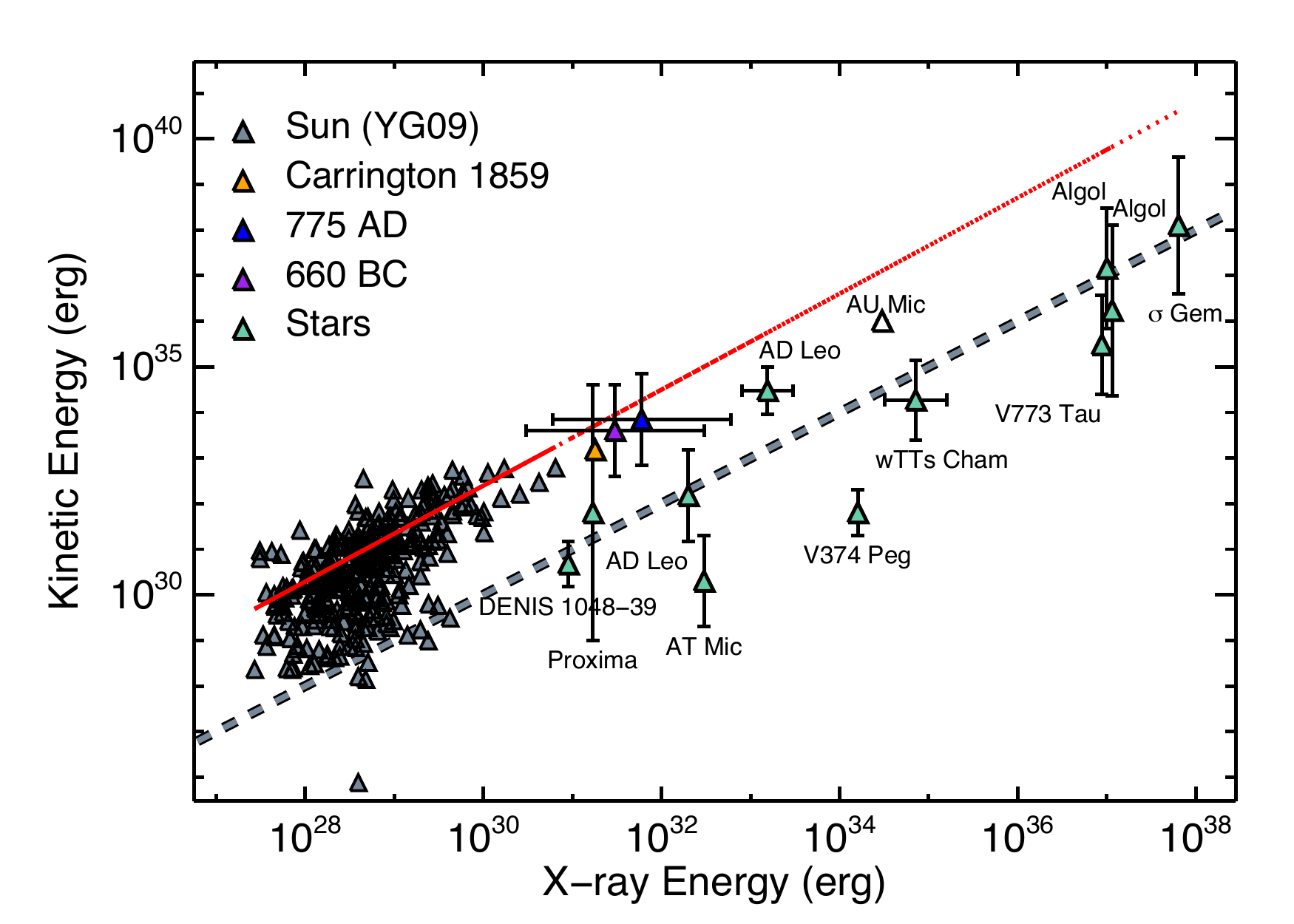}
\caption{Derived CME masses (top) and kinetic energies (bottom) as a function of the associated flare total X-ray radiated energy in the GOES 1--8~\AA\ band for solar events and the stellar events analysed here and by  \cite{Moschou:17}.  
Solar events are from the compilation of \citet[][YG09]{Yashiro2009} and are indicated by filled grey symbols. The Carrington event and the historic 775~AD and 660~BC events proposed by \citet{Melott.Thomas:12} and \citet{OHare.etal:18}, respectively, are also indicated (see text for details). Individual stellar events indicated by green filled symbols are labelled with the identity of the host star. The point corresponding to the AU~Mic event is shown hollow to designate its more doubtful CME status (see text). The red lines represent the fit to the solar data by \citet{Drake.etal:13}.  The grey dashed line in the upper panel represents a constant ratio of CME mass to flare X-ray energy.  In the lower panel, the grey dashed line represents parity between flare X-ray and CME kinetic energies \citep[see][for further details]{Drake.etal:13} } 
\label{fig:solar}
\end{center}
\end{figure}

{\subsection{660 BC event}}
{The 775 AD event, shown in Fig.~\ref{fig:solar} and further discussed in \citet{Moschou:17}, is not the only case where significant enhancements of proton fluxes were inferred by studying radionuclides in ice cores. Recently, \citet{OHare.etal:18} estimated the proton fluence for an event dated to 660 BC. The 660 BC solar proton event was found to be one order of magnitude stronger than any solar event recorded during the instrumental period and of the same order of magnitude as the 775 AD event. \citet{OHare.etal:18} estimated proton fluences of $F_{30}=2.1\times10^{10}$ proton~cm$^{-2}$, $F_{100}=6.3\times10^{9}$ proton~cm$^{-2}$, $F_{360}=1.6\times10^{9}$ proton~cm$^{-2}$ for protons with more than 30, 100, and 360 MeV fluencess respectively.} 

{Following a similar method as presented in \citet{Cliver.etal:14} and using available scalings from solar data we provide the best guess for the characteristics of an associated CME and flare. First, using the broad scatter relation derived by \citet{Cliver.Dietrich:13} we find a best guess for the soft X-ray GOES 1-8~\AA\ equivalent fluence to the $F_{30}$ proton flux of the 660 BC event of the order of $\sim20$~J~m$^{-2}$. 
The emitted soft X-ray energy is then $3\times10^{31}$~erg, which corresponds to a flare bolometric energy of $1.25\times10^{33}$~erg, according to Figure~3 in \citet{Cliver.Dietrich:13}. Then, using the scatter relation of CME properties as a function of associated flare X-ray fluence derived by \citet{Yashiro2009}, we infer an associated CME mass of $8\times10^{16}$~g and kinetic energy $4\times10^{33}$~erg.}

{Here, we haved assumed a semi-opening angle of 90$^\circ$ for the CME based on solar statistical studies for energetic X-class flares \citep{Yashiro2009,Aarnio2011SoPh}, as explained in Section~\ref{sec:analysis} and thoroughly discussed in \citet{Moschou:17}. This is significantly larger than the 24$^\circ$ opening angle assumed by \citet{Melott.Thomas:12} for the 775~AD event.  We take this opportunity to update the CME and flare parameters for the 775~AD event to account for the larger, more likely, opening angle. The resulting characteristics are  $E_k\approx7\times10^{33}$~erg for the CME kinetic energy, $2\times10^{17}$~g for the mass, $2.2\times10^{33}$ for the flare bolometric energy and $6\times10^{31}$~erg for the flare energy emitted in soft X-rays. }

{We note that for these analyses we rely on multiple scaling relations based on data with significant scatter  \citep{Yashiro2009,Cliver.Dietrich:13}. As a result, different papers in the literature have inferred different kinetic energies for the 775 AD event, from $2\times10^{33}$~erg in \citet{Melott.Thomas:12}, to $2\times10^{35}$~erg in \citet{Miyake.etal:12}, and an intermediate value $3\times10^{34}$~erg as computed in \citet{Cliver.etal:14}. 
For this reason, in Figure~\ref{fig:solar} we have added error bars of a factor of 10 reflecting the scatter in Figure~15 of \citet{Cliver.Dietrich:13}, which we used to infer the soft X-ray flaring energy. Using a factor 10 for the X-ray flaring energy then translates into a similar factor $\sim$10 uncertainty for the inferred kinetic energy of the CME \citep[see Fig. 3 and discussion in][]{Cliver.Dietrich:13}. For the CME mass we also use an error of a factor of 10 based on the spread of the solar data in \citet{Yashiro2009}.}

{\subsection{AU Mic, July 1992}}

{
\citet{Cully.etal:94} observed an EUV superflare with energy $3\times 10^{35}$~erg on AU Mic using EUVE in July 1992. Even though no direct evidence of a Doppler shift was found, a rapidly expanding CME was considered to explain the flare decay with a mass of $\sim 10^{20}$~g, and a kinetic CME energy of $10^{36}$ erg, i.e.\ 4--5 orders of magnitude higher than solar events. AU Mic has an escape speed of $375 \mathrm{km\ s^{-1}}$, since $M_*=0.31M_\odot$ and $R_*=0.84R_\odot$. The mass and kinetic energy estimated therein corresponds to a CME speed of $1,400 \mathrm{km\ s^{-1}}$, which is larger than the AU Mic local escape speed. Later, \citet{Katsova.etal:99} argued that the CME explanation is improbable given the lack of Doppler shifts or line broadening, that evidence pointed toward high plasma density, possibly reaching $10^{12}$~cm$^{-3}$ during the flare decay rather than low densities expected of an expanding CME, and finding that adiabatic expansion alone would give a slower flare decay than observed.}

\section{Results}
\label{sec:results}

Using the stellar CME candidates described above we estimated the mass and kinetic energy of each event assuming that they escaped the gravitational potential of the host star. Our results are presented in Table~\ref{tab:cme}. For the events observed through Doppler-shifts we show the waveband in column 6 and the observed fluence in that waveband in column 7. Then using the statistical relation (\ref{eq:empirical}) we convert the observed fluences to soft X-ray fluence. The observed CME plasma speeds are presented in column 3 and the estimated masses in column 4. Based on these values we estimate the CME kinetic energy (5th column) for the cases where this estimation was not performed on the original paper. 

For the events observed through X-ray absorption we have chosen an e-folding time (6th column) for the column density decay and two length scales (7th column), one obscuring (equal to the flaring loop size $L_{obs}=L_{flaring\ loop}$) and one dynamic length scale ($L_{dyn}=5R_\star$). Based on these length and time-scales we then calculate the CME speed (3rd column), mass (4th column) and kinetic energy (5th column). The last (8th) column indicates the X-ray fluence for each case in the GOES passband (1\,--\,8\AA). The conversions from each X-ray instrument to GOES fluences were performed using the \textit{MEKAL} plasma model.

The analyzable CME candidates presented in Table~\ref{tab:cme} are plotted in Figure~\ref{fig:solar} together with typical solar events and historic energetic events. Our data size is not large (only 12 points) and the event characteristics have large errors due to the indirect way of inferring them. However, we can already draw some interesting conclusions.

\paragraph{CME masses:} The top panel of Figure~\ref{fig:solar} shows the estimated stellar CME masses \citep[present work and ][]{Moschou:17} with green square symbols, overplotted with solar events in grey squares from the \citet{Yashiro2009} compilation \citep[see also][]{Drake.etal:13} and three historic energetic solar events with distinctly colored squares (Carrington event, 775 AD event \citep{Melott.Thomas:12}, and 660 BC event \citep{OHare.etal:18}). The inferred characteristics of the stellar events appear to follow the extrapolated solar trend from \citet{Drake.etal:13}, albeit with a reasonably large spread of points and uncertainties in the stellar data.  We return to this relation in Sect.~\ref{s:flarevscme} below.

\paragraph{CME kinetic energies:} We illustrate in the bottom panel of Figure~\ref{fig:solar} the estimated stellar CME kinetic energies \citep[present work and ][]{Moschou:17} with green triangles, overplotted with solar events in grey triangles \citep{Yashiro2009,Drake.etal:13} and historic energetic solar events with distinctly colored triangles of distinct color(Carrington event, 775 AD event \citep{Melott.Thomas:12}, and 660 BC event \citep{OHare.etal:18}). The kinetic energies of energetic stellar events appear to deviate from the extrapolated solar trend of \citet{Drake.etal:13}. 
This is consistent with 
the conclusions drawn in \citet[][]{Drake.etal:13} that the solar CME-flare relation is most probably breaking down in the very active stellar regime and simple solar extrapolation overestimates the fraction of the bolometric energy that a stellar CME takes from an active star (see also Sect.~\ref{s:flarevscme} below).

\section{Discussion}
\label{sec:discussion}

\subsection{Sources of Error and Discrepancy}

The analysis of historical CME candidates presented here necessarily involves a number of approximations and assumptions that inevitably lead to non-negligible sources of systematic uncertainty.  We discuss some aspects of these uncertainties below.

Emission from stellar flares and the emitting plasma characteristics inferred can be used to study stellar CMEs assuming that the solar CME - flare relation can be extended in the stellar regime \citep[e.g.][]{Aarnio.etal:12,Drake.etal:13}.  
However, flaring properties between the Sun and active stars may differ substantially.
As \citet{Briggs.Pye:03} argue, the quiescent main mass of the solar coronal plasma emits X-rays corresponding to 1-2 MK with flaring plasma reaching temperatures of the order of 10 MK. While active stars have coronae with plasma in the 20 MK range and flaring plasma corresponding to 100 MK.
In the solar corona, low first ionization potential (FIP) elements, e.g. Mg, Si, Fe, are generally overabundant in comparison to high FIP elements such as O, Ne and Ar. In contrast, in very active stellar coronae the pattern is reversed into an ``Inverse FIP Effect'', with low FIP elements appearing underabundant relative to high FIP elements. These abundance fractionation patterns appear to be a function of both activity level and spectral type \citep[see, e.g.,][]{Drake:03,Robrade.Schmitt:05,Laming:15,Wood.etal:18}.
This is evidence that stellar CMEs could consist of plasma with different characteristics---both temperature and chemical composition---than the plasma in solar CMEs. In order to better comprehend the stellar CME-flare relation it is important to combine both observational and computational work.

It is not a straightforward task to discriminate between escaping atmospheric material and photospheric evaporation. In most cases spatial resolution is a problem when trying to locate stellar flares, which is important for flaring loop length estimations, since most stars appear as point sources from the Earth's orbit. More specifically, \cite{Covino.etal:01} note that most flaring loop size estimations have been performed based on theoretical models, which make assumption for the heating during the decay phase \citep[see, e.g.][for a sustained heating]{Reale.etal:97}, apart from the unique case of \citet{Favata1999}, where the flare was pinpointed to originate from the south pole of Algol B. \citet{Covino.etal:01} argue that errors can arise through this process, especially when heating is present during the flare decay phase. Things get even more complicated when we consider that often times there are two or more plasma temperature components present during the evolution of a flaring event \citep[see, e.g.][]{Covino.etal:01}.

It is not only the plasma temperature, but also the metallicity and column density that will play a role in the relevant physics. When there is a high enough count-rate and observed spectra can be time-resolved one can fit the observations with plasma models and infer these quantities. For example \citet[][]{Covino.etal:01} used different 1T, 2T, and 3T temperature models to fit their observations and were able to derive a global coronal metallicity Z for the active star Gl 355 (LQ Hya) that is of the order of $\mathrm{Z/Z_\odot}\sim0.1$, indicating that the corona is characterized by the ``Inverse First Ionization Potential'' effect referred to above. 
This serves as an extra pointer that the coronal conditions in active-stars might differ substantially and simple extrapolations from solar events should be treated with some caution.


\subsubsection{CMEs inferred using Doppler-shifts}

\begin{figure}[htbp]
\begin{center}
\includegraphics[trim={0cm 0cm 0cm 0cm},clip, width=\columnwidth]{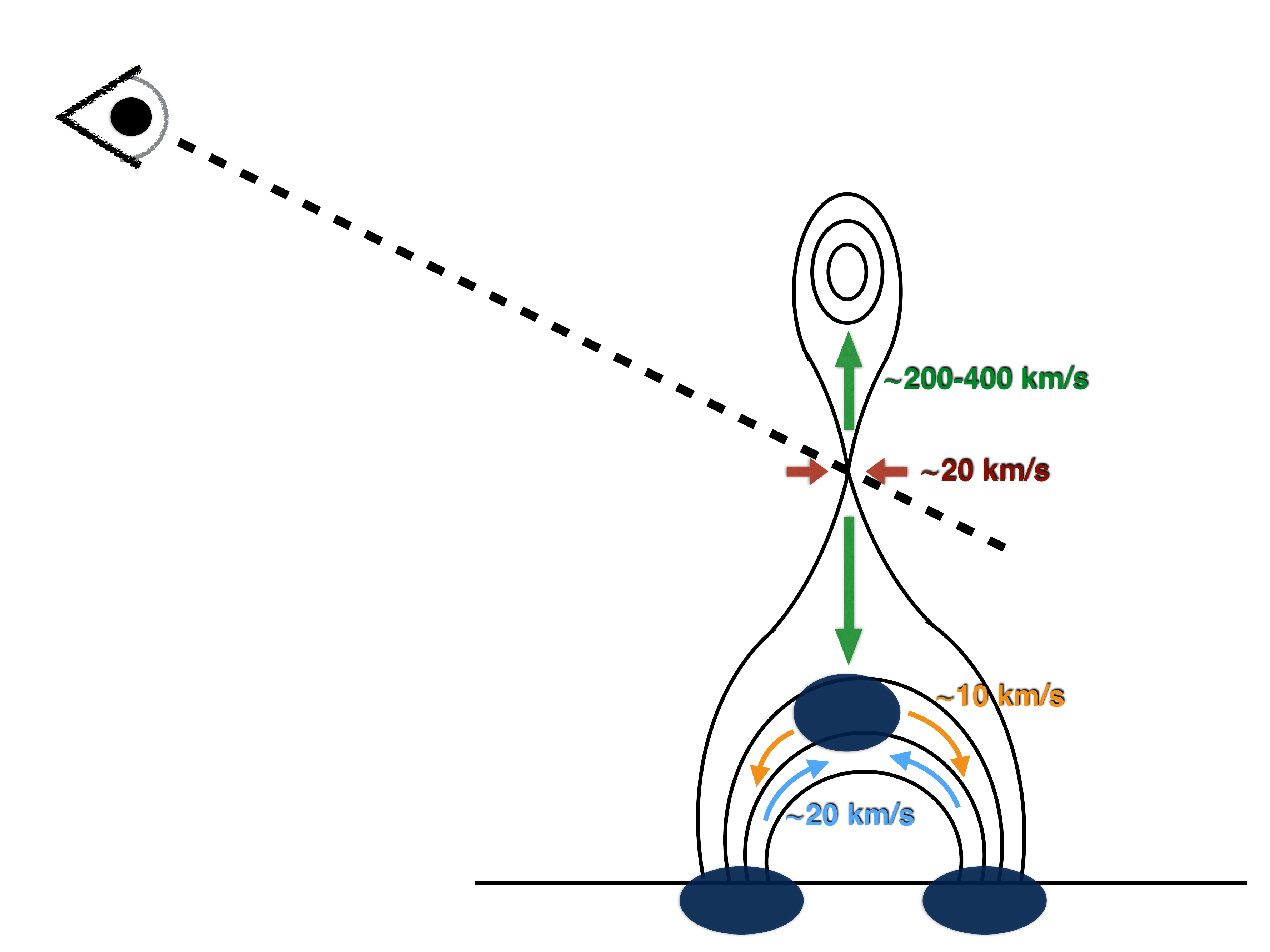}
\caption{Schematic demonstrating the standard CME-flare scenario in the Sun and the multiple flows triggered by reconnection and the subsequent atmospheric response according to \citet{Hara.etal:11}. The blue blobs indicate the X-ray emission sources.}
\label{fig:rec}
\end{center}
\end{figure}

Blue- and red- shift signals can arise from plasma motion that does not necessarily arise from  escaping outflows that are associated with CMEs.  In Figure~\ref{fig:rec}, we illustrate the scenario of a reconnection-triggered flare and all the different up- and down-flows generated as a result. Depending on the line profiles used to estimate the Doppler-shift, different atmospheric layers are probed.

Blueshifts in the range of a few tens to a couple hundred $\mathrm{km\ s^{-1}}$ that have often been reported can be confused with chromospheric evaporation, which is essentially the response of the chromosphere when accelerated particles from the corona collide with the dense local material and heat it up causing evaporation. This evaporation is an upflow, but the material is generally contained within closed magnetic loops and does not escape.

\begin{figure}[htbp]
\begin{center}
\includegraphics[trim={7cm 0cm 9cm 1cm},clip, width=\columnwidth]{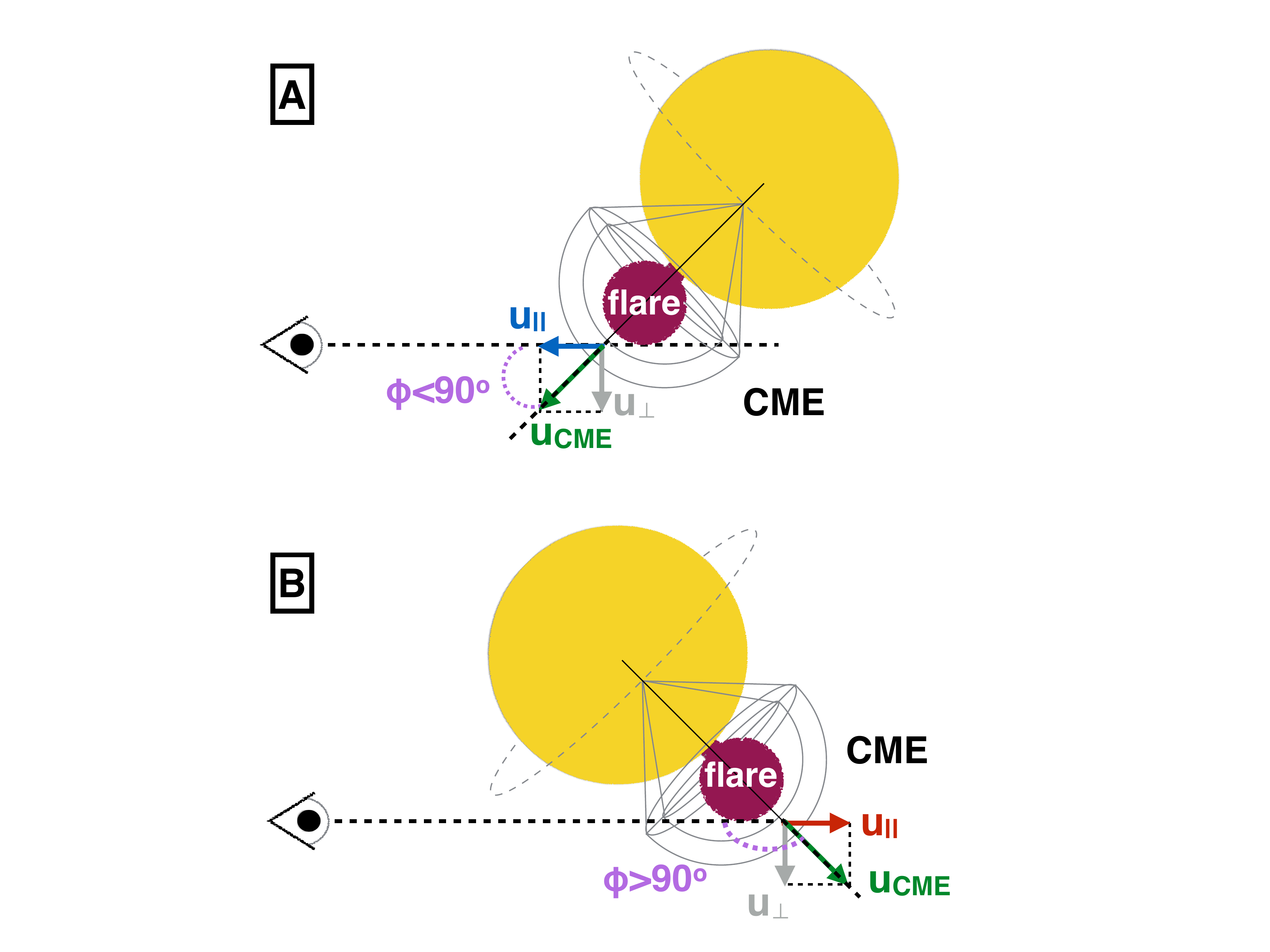}
\caption{Schematic demonstrating the scenario of a CME expanding and propagating away from the star and giving rise to either blue or red shift signatures as a result.}
\label{fig:redshift}
\end{center}
\end{figure}

\citet{Bopp.Moffett:73} estimated {\em red} shifts of the order of 1100 $\mathrm{km\ s^{-1}}$ in Balmer lines and 600 $\mathrm{km\ s^{-1}}$ in the Ca II K line for UV~Ceti.
Strong redshifts might arise from material falling onto the stellar surface but could also be associated with CMEs that are traveling in the opposite hemisphere and direction to that facing the observer; see the second panel of Figure~\ref{fig:redshift}. Blue shift measurements of speeds larger than the local escape speed thus provide a more conclusive signature of material escaping the gravitational pull of the star than red shifts, which might or might not indicate escaping material. 

As \citet{Houdebine.etal:90} mention, projection effects can be very important and add an extra level of uncertainty in the measured speeds. More specifically, projection effects allow only for the determination of lower limits of the true CME speeds, which could be severely underestimated if the CME propagation direction forms a large angle with the line-of-sight. This concept is demonstrated in the first panel of Figure~\ref{fig:redshift}. 

Another point of confusion could arise from the fact that CMEs and monster CMEs from active stars \citep{Moschou:17,Alvarado.etal:18} are large structures comprising an array of heterogeneous plasma elements that in principle could move in different directions, e.g. in halo CMEs or CMEs with large opening angles. As a result of this process, both blue- and red- shifts could be observed from the same CME event.

\subsubsection{CMEs inferred using X-rays}
Both non-eruptive and eruptive prominences are  observed frequently in the Sun. In a few stellar CME candidates \citep[e.g.][]{Ottmann.Schmitt:96}, a substantial column density increase is observed, but there is no gradual decay seen during the observation. Those cases are likely not CME events, but rather prominences that do not appear to erupt during the observation time. 
\citet{Gopalswamy.etal:03} revealed a close relation between eruptive prominences and CMEs,  determining an association rate of 83\% using microwave data. However, the \citet{Gopalswamy.etal:03} results contradict the poor association (10-30\%) found earlier by \citet{Wang.Goode:98} and \citet{Yang.Wang:02}.
More recently, \citet{Loboda.Bogachev:15} showed that most prominences (92\%) are stable and do not exhibit any apparent bulk motion. Furthermore, smaller prominences in their sample dating from between 2008 and 2009 appeared to be more dynamic than larger ones, with eruptive prominences following the same trends with the only difference being that they were larger. Thus, \citet{Loboda.Bogachev:15} concluded that there is a critical prominence mass beyond which further mass-loading will lead to eruption. In other words, only massive enough prominences will erupt.

It is worth noting that most CME candidates found in X-ray observations had inferred loop flaring lengths much larger than solar events \citep{Covino.etal:01}, often times with semi-lengths of loops larger than the stellar radius. Furthermore, \citet{Covino.etal:01} argue that large flaring loop sizes are required to account for the large flaring energy without unrealistically high magnetic fields. \citet{Covino.etal:01} then went a step further and compared the estimation of flaring loop sizes between a hydrodynamic decay model accounting for a sustained heating often reported during the decay phase \citet{Reale.etal:97} and the order of magnitude estimation presented in \citet{Pallavicini.etal:90}, only to conclude that the results in terms of flaring volume and densities were not too dissimilar (less than a factor 2 difference in all length estimations).

Later on, \citet[][]{Reale.etal:05} used nanoflares to heat coronal loops in MHD models. Heat pulses produced due to nanoflares finally heat the loop up to 1 - 1.5 MK. 
More recently, \citet{Reale:16} used MHD models to show that there are large amplitude oscillations in flare light curves if the nanoflare heat pulse is faster than the sound crossing time of the emitting loop. \citet[][]{Reale:16} explained that this takes place as there is not enough time for pressure equilibrium to be reached during the heating phase and shock waves are formed. Based on the fact that these oscillations are characteristic and differ from classic MHD waves, \citet[][]{Reale:16} was able to develop a new diagnostic for observing non-flaring coronal loops in both the solar and stellar regimes.

\subsection{Other Nominal CME Observational Methods}

\subsubsection{Type II Radio Bursts}
The comparative rarity of the CME candidates investigated in this study---only fifteen events from the last several decades---is a testament the difficulty in detecting and observing them. Until synoptic observations of the X-ray sky can be made with with sensitivity approaching that of current observatories, the number of X-ray absorption CME candidates is going to remain very low.  Similarly,continuous wide-field spectroscopic monitoring will be required to significantly increase the rate of acquisition of Doppler shift events. 

Type II radio bursts are instead the most promising observational method for CME tracking, due to their 1:1 association with CMEs in the solar case.  More than 120 hours of observations with the \textit{VLA}\footnote{The Very Large Array (VLA) is an interferometer array, using the combined views of its 27 antennas to mimic the view of a telescope as big across as the farthest distance between its antennas, i.e. 22 miles (\url{https://public.nrao.edu/telescopes/vla/}).} have been invested in the search for stellar CMEs in active stars \citep{Crosley.etal:16,Crosley.Osten:18a,Crosley.Osten:18b,Villadsen.Hallinan:18}, without any Type II radio burst detection yet, however.

Scintillation of background radio sources could also be a potential method for observing stellar CMEs as mentioned in \citet{Osten.Wolk:17}. Interplanetary scintillation has already been applied to solar CMEs \citep[e.g.][]{Manoharan:10}. \citet{Manoharan:10} used measurements of scintillation for a large number of radio sources based on which he was then able to reconstruct a three dimensional view of a propagating CME events.

The near future {\it Square Kilometer Array} will offer the potential to be able to detect large CMEs on nearby stars.  Until the commissioning of that facility, we do not envisage that the stellar CME sample presented here will be greatly enlarged upon.

\subsubsection{EUV Dimmings}
In the solar case, when a CME erupts and propagates away from the Sun, it vacates the low lying solar atmospheric material inside the CME base. This process leaves a well observed footprint in EUV wavelengths known as EUV dimming \citep[see, e.g.][]{Zhukov.Auchere:04,Mason.etal:14,Chandra.etal:16}. The most widely adopted interpretation of the coronal dimming signature is due plasma evacuation resulting from an escaping CME. However, it is also possible that the coronal material changes its temperature (thus becoming dimmer or darker in filtergrams such as the ones from SDO/AIA), while largely remaining in its original volume.

Evidence of a stellar UV dimming event on EV Lacertae was reported by \citet{Ambruster.etal:86}. EV Lac was observed over a timespan of 9 days for 4 hours per day by \textit{IUE}. \citet{Ambruster.etal:86} noted a 1.5 hour dimming of some UV wavelengths. Specifically, prominent UV line fluxes (C IV and Mg II) dropped by a factor $\sim$ 2 for about 1.5 hours. \citet{Ambruster.etal:86} favored the scenario of a large coronal mass ejection as an explanation of their results.

There is currently no EUV observing facility flying with which to investigate dimming events on stars other than the Sun.  Coronal dimming on the Sun amounts to only a few percent of the total signal \citep{Mason.etal:16}, and is aided by spatial resolution of the solar disk. EUV dimming detections will be challenging for stars and long, continuous observations will be needed, should a future EUV-capable observatory become available.

Finally, as \citet{Osten.Wolk:17} note that other potential methods to observe stellar CMEs could be a) pre-flare ``dips'' like the one reported in \citet{Giampapa.etal:82} for EQ Peg and b) the observational effect of CMEs in their immediate stellar surroundings, by for example sweeping debris disks away from the host star \citep[see, e.g.][]{Melis.etal:12,Osten.etal:13}. Interestingly, \citet{Melis.etal:12} estimated that a CME event with mass of the order of $\sim 10^{20}$~g (which is well within the CME masses we estimated for stellar events, see Figure~\ref{fig:solar}) would suffice to remove debris disk material with mass $\sim 10^{21}$~g.

\subsection{White-light \textit{versus} X-ray energy partition}

In the solar case the white light component of the flare dominates over the soft X-ray emission in the GOES passband (1\,--\,8\AA) by a factor of approximately 100 \citep[e.g.][]{Kretzschmar:11,Emslie.etal:12}.  
For both Sun-like and active stars, \citet[][]{Butler:93} and \citet[][]{Martinez-Arnaiz.etal:11} showed that the X-ray emission is tens of times larger than the flux in individual Balmer lines. More specifically, \citet[][]{Martinez-Arnaiz.etal:11} showed that for Sun-like stars and lower fluences there is an equipartition between X-rays and Balmer line emission, while the X-rays become dominant and 10 times larger in the higher energies. There is currently no extensive stellar statistical study associating the white-light to X-ray emission. We can then hypothesize that if the Balmer to X-ray relation extends to the integrated white-light part of the spectrum, we expect that the energy emitted in X-rays for more active stars will increase to more than 1\%.
 However, \citet[][]{Prochazka.etal:18} for example, used sophisticated particle beam models to investigate the energy partition during the solar flare event of 2014 June 11 and specifically understand the suppressed Balmer line emission. Therein the authors showed that during the impulsive phase of the flare only the $H_\alpha$ line was in emission, while higher Balmer lines remained in absorption. This is an example of how complex the whole image can get once we get a closer look to the multi-wavelength energy distribution in flares even in the solar case.


\subsection{Flare \textit{versus} CME occurrence frequency and the nature of CMEs on active stars}
\label{s:flarevscme}

The standard paradigm for the study of stellar flares has been the temporal correspondence of the continuum optical emission and signatures in optical, radio and X-ray wavelengths due to the collisions of accelerated particles on the lower atmosphere known as the ``Neupert effect'' first established by \citet[][]{Neupert:68}. This picture is essentially that illustrated in Figure~\ref{fig:rec} that is also thought to give rise to CMEs. Were this always the case, the flare-CME correspondence would be straightforward and inference of CME occurrence could be reliably deduced from flare observations. 

An apparent breakdown of the Neupert effect in stellar flares was found by \cite{Osten:05} in a multiwavelength study of flares on EV Lac. No observable X-ray enhancement was seen for salient $U$-band flares, and the reverse was also the case. There is indeed a history of observations of a large fraction of flares on the Sun that do not conform to the standard two-ribbon flare model in which soft X-rays arise from material evaporated from the chromosphere by energetic particle beams \citep[e.g.][]{Feldman:90,Veronig:02,Fletcher.etal:11}. 

While solar observations do  demonstrate a strong association rate between CMEs and flares \citep[e.g.,][]{Vrsnak2004,Gopalswamy2010}, far from all solar flares are associated with CMEs. The association rate is observed to grow from a few percent for weak flares to 90\%\ or more for strong (X-class) flares \citep{Gopalswamy.etal:09}.  \citet{Compagnino.etal:17} showed that the CME\,--\,flare association rate also strongly depends on the temporal window considered between the onset of the flare and the identification of the CME. Depending on the duration of that window the association rate can be significantly lower than a 1:1 association rate, even for  X-class solar flares. 

In the stellar regime, while numerous stellar flares and superflares are observed in Sun-like and active stars, there is a clear observational discrepancy between stellar flares and CMEs: as we have emphasized here, stellar CMEs cannot be imaged directly with  current technological capabilities and their indirect detection is currently extremely difficult. Existing studies extrapolating flare--CME relations for the Sun have assumed flare--CME association rates are the same.

\citet{Aarnio.etal:12} estimated stellar CME frequencies using solar extrapolations for T Tauri stars. \citet{Drake.etal:13} used solar CME\,--\,flare relations and measured solar CME properties to estimate the mass and kinetic energy stripped away from stars as a function of X-ray activity level \citep[see also this approach applied by][]{Osten:15}.  \citet[][]{Drake.etal:13} noted that the inferred mass loss and energy loss rates were unrealistically high for the most active stars. Kinetic energy requirements were especially problematic, amounting to about 10\%\ of the stellar bolometric energy output. \citet[][]{Drake.etal:13} concluded that either the relationships between solar CME mass and speed and flare X-ray energy must break down for the most active stars, or else the flare-CME association rate must drop significantly below 1.

The stellar CME results in Figure~\ref{fig:solar} present a potential way out of the \citet{Drake.etal:13} quandry. As noted in Section~\ref{sec:results}, the stellar CME kinetic energies vs.\ flare X-ray energies lie significantly {\it below} the extrapolated solar relation.  The inferred energies of the stellar events are in fact either at or beneath the line of parity drawn at equal kinetic and X-ray energy and placed about a factor of 200 lower than the extrapolated relation. This result is consistent with the \citet{Vida.etal:19} analysis, wherein scientists suggest that the event masses increase faster than event speeds (see also Section \ref{sec:analysis} and Eq. \ref{eq:vida})). The implication is that kinetic energies of large stellar CMEs are more than two orders of magnitude less energetic than extrapolation suggests. Instead of CMEs on the most active stars requiring 10\%\ of the bolometric stellar energy output, these events then suggest that less than 0.1\%\ is needed, which is then comparable to the X-ray luminosity at activity saturation, $L_X/L_{bol}=10^{-3}$.  In short, if our derived CME energies are correct, the problematic energetic requirements noted by \citet{Drake.etal:13} are alleviated. 

At the same time, our inferred CME masses are more or less in agreement with the \citet{Drake.etal:13} extrapolation and the total CME mass loss estimates by those authors still hold. For the most active Sun-like star at a saturated $L_X/L_{bol}=10^{-3}$ activity level, the \citet{Drake.etal:13} extrapolation implies a mass loss rate of a few $10^{-10}M_\odot$~yr$^{-1}$, and falling off with decreasing X-ray luminosity according to $\dot{M} \propto L_X^{1.5}$.  This can be compared with the current solar wind mass loss rate of approximately $2\times 10^{-14}M_\odot$~yr$^{-1}$ \citep[e.g.,][and references therein]{Cohen:11}. 

The ``kinetic energy deficit'' of very large stellar CMEs compared with solar extrapolations indicates that it is the CME velocity that is limited at the very energetic extreme rather than the mass. The velocity could be somewhat self-limiting in being related both to the large scale magnetic field of the host star that can act as a restraining term \citep[e.g.,][]{Alvarado.etal:18}, and to the magnetic field it carries and its interaction with the stellar wind that provide a drag force \citep{Cargill2004,Vrsnak2004,Zic2015,Moschou:17}. All of these retarding influences get stronger with increasing magnetic activity.

The other way out of the CME energy budget conundrum noted by \citet{Drake.etal:13} is a substantial decrease in the CME--flare association rate, such that the majority of the flares we observe on active stars are {\em not} accompanied by CMEs. Indeed, the observational mismatch between CMEs and flares poses a fundamental question: whether the combination of very rare CME candidates and very common flares is purely due to observational bias, or rather whether there could be a fundamental mechanism suppressing the CME escape in specific scenarios. 

The scenario of a CME suppression mechanism was raised by \citet{Drake.etal:16} who suggested that the strong overlying magnetic field in active stars might prevent CMEs from escaping. This was examined in more detail in a recently published computational study by \citet{Alvarado.etal:18} and was also discussed by \citet{Odert.etal:17}. \citet{Alvarado.etal:18} demonstrated that a strong large-scale overlying magnetic field can suppress a CME eruption that has a poloidal magnetic flux less than twice the total flaring energy of the associated flare, i.e. $E_k \geq 2 F_{total}$, where $E_k$ is the CME kinetic energy and $F_{total}$ the total flaring energy in the entire electromagnetic spectrum. For the solar case, an overlying magnetic field of 75~G is able to suppress all the currently observable solar CME events.

Our stellar CME sample cannot address the CME-flare association rate because of the enormous observational bias toward flares rather than CMEs. Such a step requires the advent of much more sensitive methods of CME detection.

\subsection{No way out for the early faint Sun paradox}

Copious mass loss through CMEs is potentially of great importance to the
``early faint Sun paradox'' first noted by \citet{Sagan.Mullen:72}. Those authors pointed out that the lower solar luminosity earlier in the history of the solar system implies global mean Earth temperatures below the freezing point of seawater until about 2.3 Gyr ago, in contradiction with geological evidence for liquid oceans. An early Sun more massive by several percent that has since been lost through mass loss provides a potential solution \citep[e.g.,][]{Guzik.etal:87,Sackmann.Boothroyd:03}.
Unfortunately, the conclusions of \citet[][see also \citealt{Osten:15}]{Drake.etal:13} would still hold: the CME-driven mass loss rate would be insufficient for the 2~Gyr old Sun to account for the luminosity deficiency. Nevertheless, it would be of considerable interest to examine the observable consequences of very high $\sim 10^{-10}M_\odot$~yr$^{-1}$ mass loss rates for the most active stars: this regime might prove the most promising for testing CME-flare relations for the combination of frequent flaring and highest flare energies.

\subsection{Relevance for exoplanet impact}

Understanding the effects of stellar transient events on exoplanetary atmospheres is not straightforward. In two computational studies \citet{Cohen.etal:11} and \citet{Cherenkov.etal:17} simulated the interaction between CMEs and the magnetospheres of hot Jupiters.  \citet{Cohen.etal:11} found that the planetary atmosphere is mostly shielded from the transient, even for a planetary magnetic field as low as 1~G, but did not perform an atmospheric loss analysis. \citet{Cherenkov.etal:17}, however, suggested that the interaction of CMEs with the hot-Jupiter envelope would substantially increase the mass-loss rate, with faster CMEs causing a higher mass-loss rate.
A limitation in CME velocity that our results suggest is interesting from the perspective of extrapolating solar CME speeds and the impact of energetic CMEs on planets.

In the compilation of CME data by \cite{Yashiro2009}, the maximum deduced speed for a solar CME is about 3000~km~s$^{-1}$ and the maximum associated flare energy is approximately $10^{31}$~erg. The stellar events examined here reach to nearly seven orders of magnitude greater flare energy and extending of the \citet{Yashiro2009} CME speed--flare energy relation this far would indicate speeds of 10,000--100,000~km~s$^{-1}$ should be reached. The stellar CME kinetic energies we infer instead indicate CME speeds should not be greatly in excess of those observed in the solar system. 
In terms of planetary impact, the ram pressure for an energetic stellar CME with density$\rho$ and moving at speed $v$, $\rho v^2$, is also expected to be two orders of magnitude lower than inferred from extrapolating solar data. 

For a dipolar planetary magnetic field, the magnetospheric standoff distance at the equator, $r_M$, scales with wind (or CME) ram pressure, $P_{ram}$, as $r_M\propto P_{ram}^{-1/6}$ \citep{Schield:69,Gombosi:04}, such that a lower pressure by a factor of 100 implies a larger magnetosphere by a factor of $\sim 2$. The implication is that the ability of CMEs to dynamically strip planetary atmospheres is much reduced compared with CME energies inferred from the \citet{Drake.etal:13} extrapolation.

\section{Conclusions}
\label{S:conclusions}
We have presented a comprehensive study of currently known historic stellar CME candidates in the literature, commenting on their relative merits in each case. We analysed each case separately to infer the CME mass and kinetic energy in order to better understand what historic events reveal about the stellar CME\,--\,flare relation and to examine similarities and differences with solar events. 
While the analysis necessarily requires some assumptions regarding CME geometries, and resulting uncertainties for individual events can exceed an order of magnitude, the large dynamic range of CME and flare properties enable useful comparisons with solar events to be made.

CME candidates observed using two methods, namely through continuous X-ray absorption and through Doppler-shifts, were investigated. Our sample of 12 events indicates that energetic stellar CMEs appear to follow the average relation between solar CME mass and X-ray flaring energy. In contrast, stellar CME energies appear to have 200 times less kinetic energy than the solar extrapolation predicts. This latter result alleviates the problematic energy requirements that otherwise result if observed stellar flares are accompanied by CMEs with kinetic energies that follow the solar extrapolation.

The kinetic energies and masses we infer for energetic stellar events indicate that CME velocities are probably limited by retardation in the large-scale stellar magnetic field and by drag in the stellar wind. Lower resulting CME kinetic energies present a much more optimistic scenario for planetary atmospheres in close proximity to active host stars, such as on planets in the habitable zones of M dwarfs.






\acknowledgments
{We thank the anonymous referee for valuable comments and suggestions.} SPM was supported by NASA Living with a Star grant number NNX16AC11G. JJD was funded by NASA contract NAS8-03060 to the {\it Chandra X-ray Center} and thanks the Director, Belinda Wilkes, for continuing advice and support. OC was supported by NASA Astrobiology Institute grant NNX15AE05G. JDAG was supported by {\it Chandra} grants AR4-15000X and GO5-16021X. CG was supported by SI Grand Challenges grant ``Lessons from Mars: Are Habitable Atmospheres on Planets around M Dwarfs Viable?''. 


\bibliographystyle{yahapj}
\bibliography{references}


\end{document}